# Modeling Convective Core Overshoot and Diffusion in Procyon Constrained by Asteroseismic Data

Short Title: Modeling Procyon

D. B. Guenther

*Department of Astronomy and Physics, Saint Mary's University, Halifax, Canada, B3H 3C3*

d_guenther@ap.smu.ca

P. Demarque

*Department of Astronomy, Yale University, New Haven, U.S.A, 06520*

M. Gruberbauer

*Department of Astronomy and Physics, Saint Mary's University, Halifax, Canada, B3H 3C3*

# ABSTRACT

We compare evolved stellar models, which match Procyon's mass and position in the HR diagram, to current ground-based asteroseismic observations. Diffusion of helium and metals along with two conventional core overshoot descriptions and the Kuhfuss nonlocal theory of convection are considered. We establish that one of the two published asteroseismic data reductions for Procyon, which mainly differ in their identification of even versus odd $l$-values, is a significantly more probable and self-consistent match to our models than the other. The most probable models according to our Bayesian analysis have evolved to just short of turnoff, still retaining a hydrogen convective core. Our most probable models include $Y$ and $Z$ diffusion and have conventional core overshoot between 0.9 and 1.5 pressure scale heights, which increases the outer radius of the convective core by between 22% to 28%, respectively. We discuss the significance of this comparatively higher than expected core overshoot amount in terms of internal mixing during evolution. The parameters of our most probable models are similar regardless of whether adiabatic or nonadiabatic model $p$-mode frequencies are compared to the observations, although, the Bayesian probabilities are greater when the nonadiabatic model frequencies are used. All the most probable models (with or without core overshoot, adiabatic or nonadiabatic model frequencies, diffusion or no diffusion, including priors for the observed HRD location and mass or not) have masses that are within one sigma of the observed mass $1.497 \pm 0.037$ M$_\odot$.

Subject Keywords: asteroseismology — convection — stars: evolution — stars: individual (Procyon) — stars: interiors — stars: variables: general

# 1. INTRODUCTION

## 1.1 Procyon

Procyon, an F5 IV-V star with a white dwarf companion, continues to play an important role in stellar evolution and stellar pulsation research. Its brightness, proximity to earth and resolved binary companion have enabled astronomers to determine accurate values for its mass, luminosity, effective temperature, and composition. From stellar model fits to Procyon's HR diagram position we know that it is near core hydrogen exhaustion, although, the precise phase is uncertain (Guenther & Demarque 1993; Kervella et al. 2004). Our models of Procyon also show it to have a very thin convective envelope and a small convective core (Guenther & Demarque 1993; Robinson et al. 2005). Because the propagation of $p$-modes is sensitive to boundaries between convective and radiative regions, Procyon is a prime target for investigating the properties of convective core overshoot and the depth of the convection zone using seismology. The $p$-mode oscillation spectrum of Procyon has been observed in radial velocity from a network of telescopes on the ground (Arentoft et al. 2008) and in luminosity from the MOST space telescope (Guenther et al. 2008), with the two sets of observations compared in (Huber et al. 2011). Although the oscillation spectra of standard models of Procyon are generally consistent with the observations there are some interesting discrepancies at low and high frequencies. These deviations suggest to us that the deep core and outermost surface layers of our standard models need additional physics, such as convective overshoot in the core and diffusion of helium and heavy elements in the envelope.

## 1.2 Convection

In the standard description of stellar interiors, the classical Schwarzschild-Ledoux criterion (Schwarzschild 1906; Ledoux 1947) defines the boundaries between regions in radiative equilibrium and regions which are unstable against convection. In regions of deep convection, the local temperature gradient exceeds the adiabatic gradient by a very small amount and the region is for all practical purposes in adiabatic equilibrium (Schwarzschild 1958).

The transition layers between regions in radiative and adiabatic equilibrium are complex. The detailed turbulent motions that occur in and around convective regions are difficult to model. Standard stellar models employ some form of mixing length description, a local model of convective energy transport, to approximate the thermodynamic structure and energy transport in these regions. The concept of the mixing length, was first introduced by Prandtl (1925) by analogy with the concept of mean free path in statistical mechanics. In this picture, the mixing length is the distance $l$ that a fluid parcel travels before dissolving into its environment. The value of $l$ is a free parameter of the theory, which is treated somewhat differently in modeling the surface convection zone and in describing convective overshoot.

In the surface convection zone, where we use the formulation of the mixing length theory (MLT) designed by Böhm-Vitense (1958) to describe the convective outer layers of cool stars, the mixing length is scaled in each layer within the convection zone to the local pressure scale height $H_p$ by the free parameter $\alpha$ so that we have $l = \alpha H_p$. The radii of sun-like stellar models are sensitively dependent on the choice of $\alpha$. In the case of the Sun, for a given stellar evolution code and model physics, the value of $\alpha$ is chosen to match the solar radius for a calibrated standard solar model (Demarque & Percy 1964; Bahcall et al. 2005; Serenelli 2010). And even though we know from a few well studied sun-like objects, such as $\alpha$ Centauri (Demarque et al. 1986; Fernandes & Neuforge 1995; Miglio & Montalbán 2005), that $\alpha$ can vary from star to star, this solar-tuned value is then commonly used for models of other stars and in constructing stellar evolutionary tracks, for which no reliable calibration is available (Yi et al. 2001; Bressan et al. 2012). Additionally, we note that the mixing length parameter depends on parameters of the model atmosphere, varying from star to star, as shown in the 2D hydrodynamical simulations of Ludwig et al. (1999) and the 3D simulations of Trampedach (2004).

The MLT provides an excellent approximation to the temperature structure and dynamics of the deep layers of the convection zone. Hydrodynamical models of the convective envelope of sun-like stars match the predictions of the MLT everywhere except near the surface where the superadiabaticity, i.e., the difference between the temperature gradient and the adiabatic temperature gradient, peaks (Chan & Sofia 1987, 1989). However, hydrodynamical simulations predict a more sharply peaked superadiabatic layer (or SAL), than the MLT (Stein & Nordlund 2000; Robinson et al. 2003, 2005; Arnett et al. 2010; Trampedach & Stein 2011; Tanner et al. 2012), a result whose validity is supported by the improved agreement between the $p$-mode frequencies of solar models based on the hydrodynamic simulations and the helioseismic observations (Rosenthal et al. 1999; Li et al. 2002).

In sun-like stars, the SAL is located near or just below the photosphere, and hydrodynamic simulations, that include the atmospheric layers, can provide a detailed description of convective overshoot and mixing at the top of the convection zone into the atmosphere (Freytag et al. 1996; Ludwig et al. 1999; Nordlund et al. 2009; Tanner et al. 2013). On the other hand, the region below the base of the convection zone cannot be simulated in the same detail, and the extent of overshoot mixing (e.g. the region of the solar tachocline), is still not well-understood. Mixing below the convection base has attracted much interest over the years because it can affect the abundances of light elements at the surface of the Sun and sun-like stars (Deliyannis & Pinsonneault 1990; Deliyannis et al. 1993). Recently, Lebreton & Goupil (2012) have presented evidence for overshoot below the convection zone of the sun-like star HD 52265 using asteroseismology. This paper includes a preliminary test of the effect of overshoot below Procyon's convective envelope in Section 5.

1.3 Core Overshoot

Regarding core overshoot, there are many discussions in the literature describing the possibility and consequences of the mixing of chemical species and the extension of the adiabatic layer beyond the predicted edge of a convective core. For its earliest development, see the work of Roxburgh (1965), Saslaw & Schwarzschild (1965), and Shaviv and Salpeter (1973). Core overshoot is a major source of uncertainty in the calculation of stellar evolutionary lifetimes. The extension of mixing above the edge of the convective core increases the amount of fuel available to the nuclear burning region, and, as a consequence, will increase the lifetime of the main sequence core burning phase (see, e.g., a discussion for a 1.5 $M_\odot$ star like Procyon by Maeder (1975)).

To model core overshoot, a standard approach taken in stellar evolution codes is to extend the region of chemical mixing by a distance $\beta\, H_P$ above the top of the convective core, where $\beta$ (the overshoot) is a free parameter to be determined by comparisons to observations. The value of $\beta$ has been estimated observationally in several ways: by inspection of star cluster color-magnitude diagrams (Prather & Demarque 1974; Maeder & Mermilliod 1981; Demarque et al. 1994): by studying eclipsing binaries (Ribas et al. 2000, Zhang 2012, Torres et al. 2014): by measuring stellar pulsation (Dupret et al. 2004; Montalbán, J. et al. 2013); by studying the mass luminosity relation for Cepheids (Cordier et al. 2003). We note that nearly all of these studies find evidence for modest overshoot with $0 < \beta < 0.2\, H_p$.

There is an additional problem in this simple approach for modeling core overshoot in stellar models. The unknown effects of turbulence at the core interface create an additional uncertainty in evaluating the temperature gradient in the overshoot layer, which must lie somewhere between the adiabatic gradient and the local radiative temperature gradient. The theoretical discussion of convective core overshoot by Zahn (1991) favors using the adiabatic gradient in the overshoot region (penetrative convection). On the other hand, Zhang & Li (2012), who addressed the problem using a semi-analytic turbulent convection model, derive solutions that differ from the Zahn solution. It is not possible at this time to perform 3D simulations with the required resolution for the core overshoot region. Evidence from 3D numerical simulations applicable to stellar envelopes performed in the anelastic approximation due to Brummel et al. (2002) show that the mixed overshoot region is closer to radiative equilibrium (overmixing) than adiabatic equilibrium (penetrative convection). The same result was found to hold for overshoot from the convection zone into the atmosphere in the radiation hydrodynamical convection simulations of Tanner et al. (2013).

Due to these uncertainties, we consider in this paper two variants for the local temperature gradient in the overshoot layer: one in which the over-mixed region retains the local radiative temperature gradient, and one in which the over-mixed region is forced to be adiabatic. These two treatments of the temperature gradient should bracket the actual situation in the overshoot layer.

In addition, we consider a non-local implementation of convective overshoot described by Kuhfuss (1986). Kuhfuss developed a time-dependent model of turbulent convection based on the hydrodynamic and continuity equations. It follows individual components of the fluid. It is well adapted to conventional stellar evolutionary modeling yielding averaged quantities in spherical shells that ultimately provide the velocity scale of convective motions. Specifically, in the Kuhfuss formulation, the boundaries or extent of the convective region is based on the extent of the non-zero velocities. There are several adjustable parameters in the theory, described in some detail by Straka et al (2005), which are fixed in the present paper by matching the theory's values of convective flux and velocity to that predicted by the MLT. The Kuhfuss formalism implemented in YREC (Yale Stellar Evolution Code, Demarque et al. 2008), described in Straka et al. (2005) and used in the present paper, also assumes adiabatic penetration in the overshoot region.

### 1.4 Diffusion

Our principal analysis is done for models that do not include chemical diffusion of helium and heavier elements because our implementation of diffusion fails when the mass in the convective envelope falls significantly below $\sim 10^{-5}$ $M_\odot$. The convective envelope mass of models of Procyon does fall two orders of magnitude below this threshold during a portion of Procyon's evolution off of the main sequence. The computed rate of diffusion is so high during this time that nearly all the helium and metals are drained out of the surface convection zone over a few time steps. This is probably unrealistic since no zero metallicity F stars are known. We note, also, that other mechanisms not included in our models, such as rotation, winds, radiative pressure, and turbulence at the convective base, could inhibit or reverse the effect. To complement our main analysis we have, though, constructed a grid of conventional core overshoot models that do include diffusion (see section 3). The diffusion computation itself, though, is turned off whenever the convective envelope mass drops below $2.0 \times 10^{-5}$ $M_\odot$. Although the models are certainly not completely correct, they do provide us with some idea of how sensitive our results are to the effects of diffusion.

### 1.5 Bayesian Approach and Modeling

In order to compare the *p*-mode oscillation spectrum of our stellar models of Procyon to observations we use a Bayesian approach (Jeffreys 1961). Bayesian methodology enables us to unambiguously define the modeling hypotheses (in our case, the three different models of convective core overshoot and whether or not diffusion is included), the constraining data (the *p*-mode frequencies, mass, and HR-diagram location), and any additional biases or prior assumptions (surface effects and mode bumping) and obtain a unique probability measure (the evidences and posterior probabilities) of the viability of each hypothesis with which it can then be compared. The Bayesian approach formally requires us to explicitly state our prior probabilities. We use this to our advantage, not only to introduce our prior knowledge, but also to test its consistency with the asteroseismic data. Rather than imposing implicit constraints on stellar parameters (or explicit non-asteroseismic terms in the likelihood calculation), we can therefore easily separate the impact of the asteroseismic observations and the effect of prior constraints on the posterior probabilities.

Our analysis is based on several extensive model grids and their adiabatic and non-adiabatic oscillation spectra that span a range of age, radius, luminosity, composition, mixing length parameter, and core overshoot parameter. The grids are searched using our Bayesian algorithm to locate the most probable models.

We have also computed evolutionary tracks of a 1.497 $M_\odot$ star from the zero-age main-sequence to Procyon's location in the HR diagram. These models are used to directly see what effect changing a single parameter (diffusion, core overshoot, and envelope overshoot) has on the structure and $p$-mode frequencies of the models (see section 5).

### 1.6 Organization

In the next section we describe the modeling parameters, assumptions and constraints. In section 3 we continue this discussion focusing on the issue of diffusion. In section 4 we describe the asteroseismic observational data that we use to constrain the models. In section 5 we describe the effects of changing individual parameters on the basic structure and p-mode frequencies of models of Procyon. In section 6, we present the Bayesian probabilities of the different core overshoot and diffusion models and determine the parameters of the most probable models for Procyon. And section 7, we summarize our results and discuss their implications.

## 2. MODELING PARAMETERS, ASSUMPTIONS, AND CONSTRAINTS

### 2.1 Non-asteroseismic Observational Constraints

We use the mass determination of Girard et al. (2000) of 1.497 ± 0.037 M$_\odot$. We note that the mass of Procyon is currently being debated in the literature (see, for example the recent overview in Liebert et al. 2013) as ground based (Gatewood & Han, 2006) and HST determinations differ. Girard and collaborators (private communication) are continuing to collect data from HST to revise the binary orbit of Procyon. Current estimates agree with the published mass (Girard et al. 2000) to within ±0.001 M$_\odot$. The diameter of Procyon has been measured directly using interferometry by Kervella et al 2004 to be 2.048±0.025 D$_\odot$ or log R/R$_\odot$ = 0.311±0.005. We assume the luminosity is given by logL/L$_\odot$ = 0.84±0.02 and the effective temperature is given by $T_{\text{eff}}$ = 6530±90K. These values encompass most published values (see discussions in Eggenberger 2005; Dogan, G. et al. 2010; Guenther & Demarque 1993). Also the luminosity is consistent with the Girard et al. (2000) mass determination.

The observed metallicity is believed to be near solar (Steffen 1985; Takeda et al. 1996; Kato, Watanabe & Sadakane 1996) but see discussion in Chiavassa et al. (2012). Chiavassa et al. note that 3D time-dependent hydrodynamical simulations can yield significant differences in metal abundances. Their simulations (and also those of Nordlund & Davins 1990) are better able to model the large fluctuations at the surface due to granulation in F type stars and account for the line shifts and observed bisector asymmetries in iron (Allende Prieto et al 2002). We do not constrain the composition, it being a free parameter of the grids, but when comparing our results to observations, we will take [Fe/H] = 0.0 ± 0.05 , which corresponds to $Z/X$ = 0.0245±0.003 for the Grevesse and Noels (1993) mixture that we use in our models. Note that the uncertainty does not include the uncertainty in the solar value. The observable parameters are listed in Table 1.

The luminosity, effective temperature, composition (*Y* and *Z*), the mixing length parameter ($\alpha$), overshoot amount ($\beta$), and the age are all free parameters spanned by the computed grids.

TABLE 1

Observational Data for Procyon

| Observable | Value | Reference |
|---|---|---|
| Mass [$M_\odot$] | 1.497±0.037 | Girard et al. (2000) |
| log L/$L_\odot$ | 0.84±0.02 | Dogan, G. et al. 2010 |
| log $T_{eff}$ [K] | 3.815±0.006 | Dogan, G. et al. 2010 |
| log R/$R_\odot$ | 0.311±0.005 | Kervella et al. 2004 |
| Z/X | 0.0245±0.003 | Allende Prieto et al 2002 |
| <$\Delta\nu_0$> [$\mu$Hz] | 54.6±1.8 | Scenario A, Bedding et al. (2010) |
| <$\Delta\nu_1$> [$\mu$Hz] | 54.9±1.6 | Scenario A, Bedding et al. (2010) |
| <$\Delta\nu_2$> [$\mu$Hz] | 54.7±2.0 | Scenario A, Bedding et al. (2010) |
| <$\delta\nu_0$> [$\mu$Hz] | 4.5±1.6 | Scenario A, Bedding et al. (2010) |

2.2 Observational Constraints, Asteroseismic

Procyon's oscillation spectrum was observed in 2007 simultaneously from eleven ground stations in radial velocity (Arentoft et al. 2008) and from the MOST satellite in luminosity (Guenther et al. 2008). The results are discussed and compared in Huber et al. (2011). We use exclusively the results from the ground based observations because, although the MOST data are consistent with the ground based observations, they are of lower quality. Here, we begin by comparing the two reductions of the RV multisite campaign labeled Scenario A and Scenario B by Bedding et al. (2010). The two data sets are primarily distinguished by their even and odd *l*-value identifications. Although Scenario A is preferred by their own Bayesian modeling tests, arguments in White et al. (2012) show that Scenario B provides a better model fit to the $\varepsilon$-$T_{\rm eff}$ trend for known stars, where $\varepsilon$ is the large spacing offset factor used in the asymptotic relationship for frequency as a function of *n* and *l*.

The large spacing, defined by,

$$\Delta\nu_l = \nu_{n,l} - \nu_{n-1,l} \qquad (1)$$

varies as a function of frequency. Averaged over the frequency range 550 µHz to 1200 µHz for scenario A, the large spacings are $\langle\Delta\nu_0\rangle = 54.6\pm1.8$ µHz, $\langle\Delta\nu_1\rangle = 54.9\pm1.6$ µHz, and $\langle\Delta\nu_2\rangle = 54.7\pm2.0$ µHz (Bedding et al. 2010), where the ± range is the standard deviation of the individual spacings about the average. The small spacing, defined by,

$$\delta\nu_0 = \nu_{n,0} - \nu_{n-1,2} \qquad (2)$$

also varies as a function of frequency. Averaged over the frequency range 550 µHz to 1200 µHz for scenario A, $\langle\delta\nu_0\rangle = 4.5\pm1.6$ µHz, again, where the ± range is the standard deviation of the individual spacings about the average. For reference the averaged spacings are listed in Table 1. Note that only the mass, luminosity, and effective temperature listed in Table 1 are used as priors in the Bayesian search for the most probable models.

## 2.3 Stellar Models

We computed four grids of models, the first three are distinguished by their core convection model and the fourth by the inclusion of helium and heavy element diffusion. The four grids are:

1. raOv grid: Evolved models based on a conventional overshoot algorithm that forces mixing to a specified pressure scale height above the convective core boundary as determined by the Schwarzschild criterion (Schwarzschild 1906) but which retains the regions radiative temperature gradient.

   2. adOv grid: Evolved models based on the same overshoot algorithm, modified to force the overshoot region to be adiabatic.

   3. Ku grid: a formulation based on the Kuhfuss (1986), nonlocal, model of convection.

   4. raOvD grid: Similar to raOv grid except that the effects of diffusion are partially included in the models.

We also have computed tuned models for Procyon in which a single parameter is perturbed to see directly how it affects the structure of the model and its oscillation spectrum. These models are discussed in section 5.

Each grid encompasses a broad range in mass, luminosity, surface temperature, composition ($Y$ and $Z$), mixing length parameter, $\alpha$, and core overshoot parameter, $\beta$. As noted in the introduction, for simplicity in notation, the mixing length and overshoot parameters for the Kuhfuss models are those that, within the theory, closely mimic the corresponding parameters in the MLT (Straka, et al. 2005). The specific range and resolution for the mass is 1.41 $M_\odot$ to 1.55 $M_\odot$ in steps of 0.02 $M_\odot$. We allowed for a very broad range in mass to see how well the data constrain the mass (and age). To see how well the helium and metal abundances can be constrained we allow the helium mass fraction $Y$ to span the range 0.26 to 0.31 in steps of 0.01 and the metal mass fraction $Z$ to span the range 0.014 to 0.026 in steps of 0.002. The mixing length parameter was set to values between 1.7 and 2.5 in steps of 0.2, and the overshoot parameter (scaled to the pressure scale height) was initially set to values between 0.0, i.e., no overshoot, to 1.0, i.e., overshoot of one pressure scale height in the MLT. After completing our analysis using these grids we saw, as will be discussed later, that we needed to extend the raOv and raOvD grid to include overshoot values beyond 1.0 so we extended these two grids up to 2.0.

We split our model grids into 11 (21 in the case of the extended raOv and raOvD) grids) subgrids each corresponding to the amount of overshoot used, from 0.0 to 1.0 (2.0 for raOv and raOvD) in steps of 0.1. We computed the Bayesian evidence for each subgrid given the asteroseismic data using our Bayesian code. We additionally included two priors in the probability calculations, one for the HRD location and one for the mass. The HRD location prior is a Gaussian prior given by the observed HRD constraints, i.e., $\log L/L_\odot = 0.84 \pm 0.02$ and $\log T_{eff} = 3.815 \pm 0.006$ (i.e., $T_{eff} = 6530 \pm 90$K). The mass prior is also a Gaussian prior given by the observed mass $M = 1.497 \pm 0.037$ $M_\odot$ (Table 1, section 2.1). The range of parameters for the grids are summarized in Table 2.

The selection of parameters, their range, and resolution was, in part, determined by our ability to compute the grids in a reasonable (~1 month each) amount of time. Adding another parameter, for example, a range of envelope overshoot parameters or increasing the resolution of any parameter, would, of course, scale the computation time up, accordingly.

TABLE 2

Grid Parameters

| Parameter | Range | Step size |
| --- | --- | --- |
| Mass [$M_\odot$] | 1.41 to 1.55 | 0.02 |
| $Y$ | 0.26 to 0.31 | 0.01 |
| $Z$ | 0.014 to 0.026 | 0.002 |
| $\alpha$ | 1.7 to 2.5 | 0.2 |
| $\beta$ (overshoot) | 0.0 to 1.0 (2.0) | 0.1 |

For each parameter an evolutionary track was computed using YJG a version of YREC (Demarque et al. 2008) that includes the calculation of the oscillation spectra of the stellar models based on routines from Guenther's nonradial, nonadiabatic stellar pulsation code (Guenther 1994). Constitutive physics include the OPAL98 (Iglesias & Rogers 1996) and Alexander & Ferguson (1994) opacity tables, as well as the Lawrence Livermore 2005 equation of state tables (Rogers 1986; Rogers et al. 1996). We use the Grevesse and Noels (1993) solar mixture of elements. Convective energy transport was modeled using either the Böhm-Vitense mixing-length theory (Böhm-Vitense 1958) or the Kuhfuss (1986) nonlocal mixing length theory. The atmosphere is implemented using Eddington gray atmosphere. Nuclear reaction cross-sections were taken from Bahcall et al. (2001) and the nuclear reaction rates from Table 21 in Bahcall & Ulrich (1988). The tracks were started on the Hayashi track (Hayashi 1961) above the birthline. When a model crosses the birthline, the age is reset to zero. Note that according to Palla and Stahler (1993) the pre-main-sequence evolution after the birthline for stars of Procyon's mass should be similar to the classical evolution we calculate, specifically, the models will go through a fully convective phase before leaving the Hayashi track.

We computed the $l = 0$, 1, 2, and 3 $p$-mode adiabatic and nonadiabatic frequencies for models falling within a large rectangle enclosing Procyon's position in the theoretical HR-diagram, defined by the boundaries, $3.79657 \leq \log T_{\text{eff}} \leq 3.83251$ and $0.67 \leq \log L/L_\odot \leq 1.03$.

Finally, for comparison purposes we computed a calibrated standard solar model (Demarque & Percy 1964; Guenther et al. 1992; Christensen-Dalsgaard 1982) using the same constitutive physics as for Procyon. Our reference standard solar model does not include diffusion nor does it include overshoot. For the standard solar model we assumed the Sun is 4.5 Gyr old, measured from the zero-age main-sequence. We obtained Z/X =0.0244 at the surface. The initial helium abundance was $Y_0 = 0.2714$. And the mixing length parameter was 1.72.

2.4 Bayesian Method

Our Bayesian based code was developed specifically to compare the oscillation spectra of stellar models to observations that includes free parameters to account for surface effects and mode bumping. A full description of the method and comparison to the $\chi^2$ method is contained in Gruberbauer et al. 2012. A more detailed example of its application to the Sun is described in Gruberbauer and Guenther (2013) and to sun-like stars observed by Kepler in Gruberbauer et al. (2013). It is extremely important to our analysis here that surface effects and mode bumping are dealt with using priors and marginalization because we do not want the problems modeling diffusion and convection in Procyon's envelope or the existence of bumped modes to interfere with our analysis of Procyon's core. Here we provide a brief summary of the assumptions we have used in our Bayesian approach.

The likelihood (a probability), which compares the model ($f_m$) and observed frequencies ($f_o$) and takes into account both random ($e$) and systematic errors ($\gamma\Delta$), is computed for each model. We determine the likelihood (see Gruberbauer et al. 2012) using,

$$f_0 - f_m = \gamma\Delta + e. \qquad (3)$$

The random errors are assumed to be independent and Gaussian. The systematic errors can account for surface effects but the form is general enough to account for other effects such as rotation and mode bumping. $\Delta$ is a free parameter restricted to values between 0 and maximum value $\Delta_{max}$ defined by the variable power law surface effect,

$$\Delta_{max} = \Delta\nu\left(\frac{f_m}{f_{max,m}}\right)^b, \qquad (4)$$

where $\Delta\nu$ is the asymptotic large frequency spacing (Tassoul 1980) of the model and $f_{max,m}$ is the frequency of the highest order in the model. Finally, $\gamma = \pm 1$, allowing for either positive or negative surface effects. The exponent *b* of the power law can take on values between 3.0 and 6.0, where *b* = 4.9 corresponds to the solar surface effect correction. The γΔ parameter is incorporated into the Bayesian analysis using a *beta* prior, which prefers smaller values of differences (between data and model) to larger ones (Gruberbauer et al. 2012). This approach is completely general, unlike the standard surface effect correction (Kjeldsen et al. 2008), and will correctly propagate uncertainties regardless of the actual power law form of the surface effect and even whether or not it exists.

The likelihoods for each frequency are combined to form a likelihood for each model (i.e., the product of the individual probabilities). The likelihoods for the models are weighted and properly renormalized by the mass and HR diagram location priors. The Bayesian method allows the weights to be correctly normalized so that we can propagate systematic errors and derive structural parameters of the models and their standard deviations. The grid is oversampled using linear interpolation up to the point where the probabilities no longer change. The likelihoods of each of the models in a grid are themselves combined to form a likelihood for the whole grid. Since properly normalized priors are used, we can then compute the average of the prior-weighted likelihoods for each grid. This is the overall Bayesian evidence for each grid. In other words, the evidence represents the overall probability of a given grid but with the probability correctly normalized by the priors so that it can be directly compared to the probabilities of other grids. In our analysis we assume that evidence ratios (called odds ratios) greater than a factor of 10, deemed "very strong" by Jeffreys (1961), represent a significant difference.

Our Bayesian analysis is restricted to testing only the hypothesis put forward. It cannot, unless specific tests are created, be used to determine the underlying physical causes and effects of the various scenarios investigated. Therefore, to more fully understand how adjustments to each parameter affect our models of Procyon we have computed models that fit Procyon's observed mass and HR diagram location but which have a single parameter varied (section 5). In this way we can easily see how each of the parameters directly affect the structure and *p*-mode frequencies of the models.

## 3. DIFFUSION

As noted in the introduction (section 1.4), the diffusion rates of metals and helium in Procyon are predicted to be very high (Turcotte et al 1998; Morel & Thévenin 2002). Unfortunately, for some specific parameters the calculated rates of diffusion (specifically, gravitational settling, Bahcall et al. 1995) are so high that the diffusion calculation becomes unstable as the high rates create an ever steepening composition gradient at the base of the convective envelope which in turn forces even higher rates, ultimately draining all the helium and heavy elements out of the convective envelope in just a few model time steps. Although there are well known abundance anomalies associated with F stars, no young star has ever been observed with a zero metal abundance. Numerical models show that both rotation and winds can inhibit diffusion (Chaboyer, et al, 1999) and any form of turbulence near the base of the convection zone can also inhibit diffusion (Charbonnel & Vauclair 1992). Morel & Thévenin (2002) consider radiative effects, i.e., photon-ion collisions, that counter the effects of gravitational settling, and, as they show, could be large enough to inhibit partly the effects of gravitational settling. Kervella et al (2004) have constructed models of Procyon that do take into account radiative pressure. Their models are not directly comparable to our models because of the different mixing length theory and atmosphere model. Although the diffused helium and heavy elements are subsequently mixed back into the convective envelope when, as the star evolves toward the giant branch, the base of the convection zone deepens, we chose to leave diffusion completely off for our comparison of the three different core overshoot models but do include it in our fourth grid raOvD so we can see how its inclusion might affect our results.

Our implementation of gravitational settling of helium and heavy elements is based on the diffusion formulation of Bahcall and Loeb (1990) and has been used successfully in solar models (e.g., Bahcall & Pinsonneault 1992; Thoul et al. 1994; Morel et al. 1997; Bahcall et al. 2001) and general post-main-sequence evolutionary calculations (e.g., Deliyannis et al. 1991; Chaboyer et al. 1992; Straniero et al. 1997; Demarque et al. 2004). At each time step, the diffusion equations are solved to obtain a revised run of composition. For most scenarios the process is relatively stable only running into difficulty when the change in composition is relatively large. With large changes to the composition, the opacities are affected enough to perturb the location of the base of the convective envelope, which, depending on the composition gradients that have developed perturbs the run of composition. The process becomes unstable because moving the convective envelope base up and down is not self-correcting, i.e., we cannot unmix what has already been mixed.

To test the behavior of our diffusion implementation for very small convective envelope masses we computed evolutionary tracks (no core overshoot) of a 1.497 $M_\odot$ star with solar composition starting from the ZAMS. We modified the diffusion calculation so that it would turn itself completely off whenever the mass of the convective envelope dropped below a threshold value, $M_D$. For a 1.497 $M_\odot$ star the convective envelope will drop just below $1\times10^{-7}$ $M_\odot$. In figure 1 we plot the mass of the convective envelope versus age for $M_D = 5\times10^{-5}$, $1\times10^{-5}$, $5\times10^{-6}$, and $2\times10^{-6}$ $M_\odot$. (We are unable to compute realistic evolutionary models for $M_D < 1\times10^{-6}$ $M_\odot$.) The mass fraction of hydrogen in the core, $X_c$ is indicated along the top axis. As soon as the envelope mass rises above the threshold value, $M_D$, and diffusion is turned on, diffusion starts perturbing the location of base of the convective envelope, inhibiting slightly its growth. In figures 2 and 3 we see the effect of diffusion on $X_s$, the mass fraction of hydrogen at the surface, and $Z_s$, the metallicity at the surface. For the smallest threshold tested, $M_D = 2\times10^{-6}$ $M_\odot$, helium is almost completely drained out of the convective envelope as soon as diffusion gets turned on. The metals are also shown to drain out of the convective envelope but for $M_D < 1\times10^{-5}$ $M_\odot$ the unstable interaction between the location of the convective envelope base and the chemical abundances produces large fluctuations in $Z_s$ (also noticeable in $X_s$).

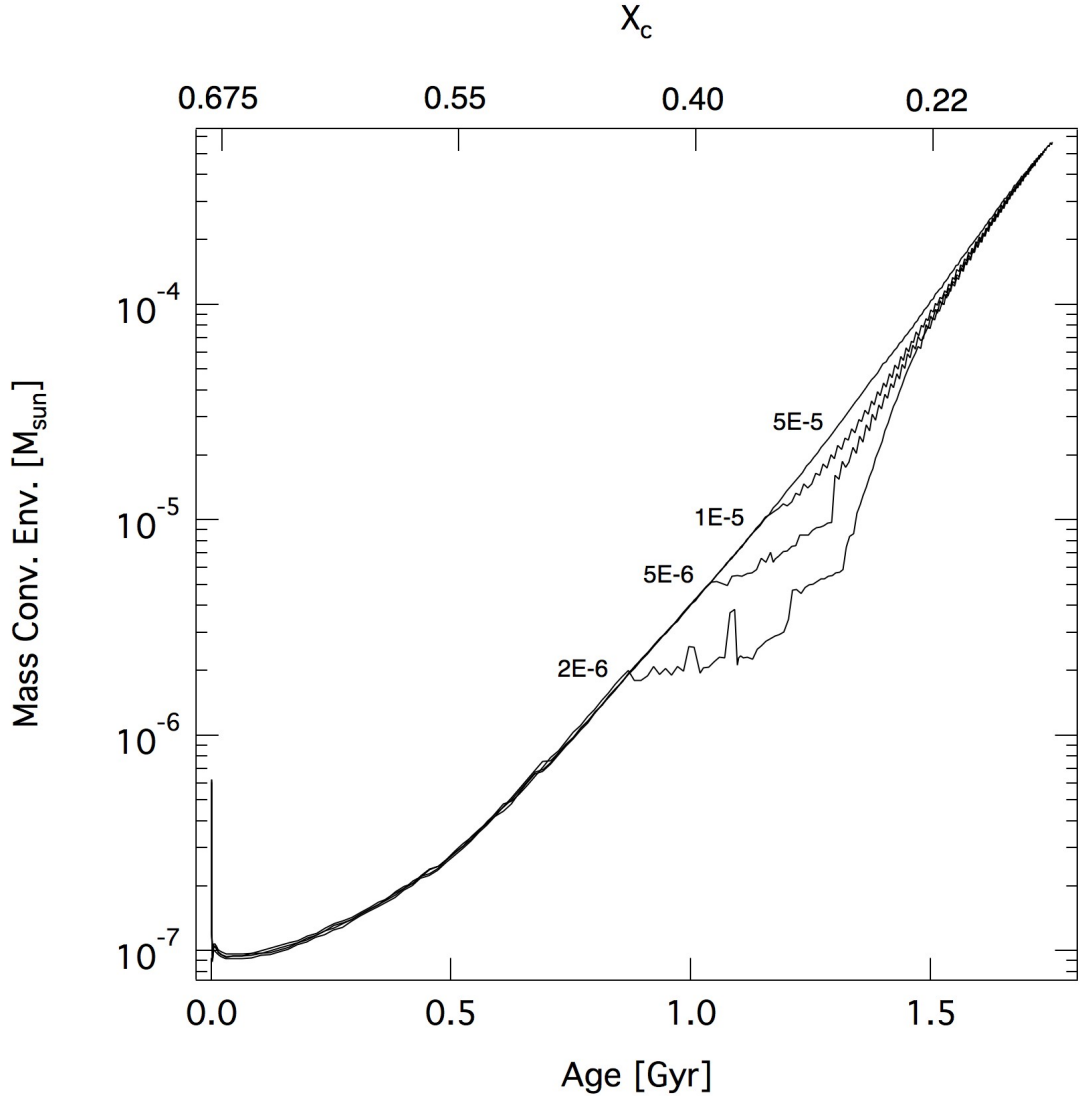

Fig. 1 — Mass of the convective envelope for $M_D = 2.0\times10^{-6}$, $5.0\times10^{-6}$, $1.0\times10^{-5}$, and $5.0\times10^{-5}$ $M_\odot$ as a function of age for a 1.5 $M_\odot$ stellar evolutionary track. The curves deviate from each other at the point where the convective envelope mass rises above $M_D$

and diffusion of $Y$ and $Z$ is turned on in the code. The top axis shows the mass fraction of hydrogen, $X_c$ in the core.

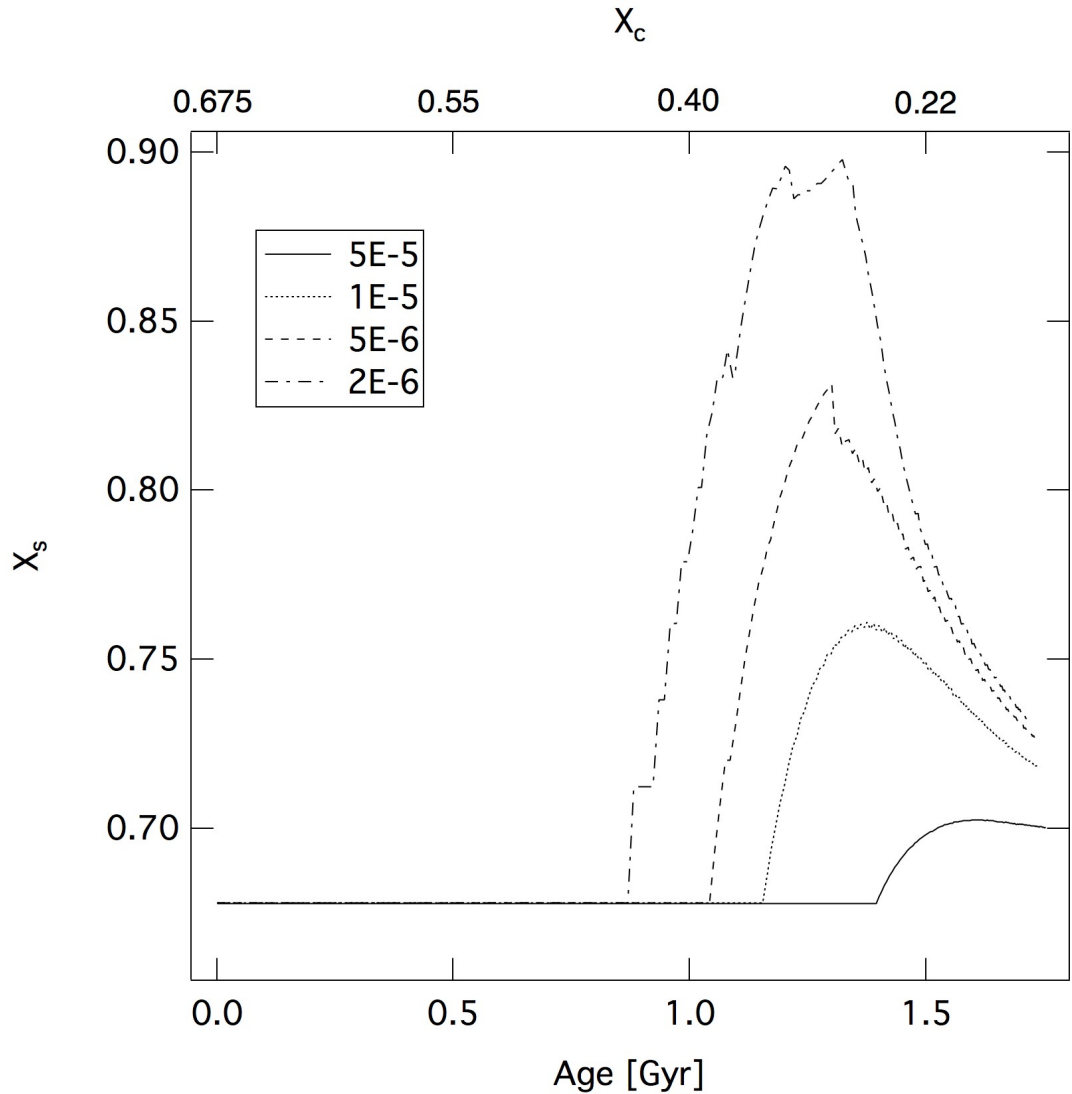

Fig. 2 — The time evolution of the surface mass fraction of hydrogen, $X_s$ for a 1.5 M$_\odot$ evolutionary track. The different curves correspond to different $M_D$, as indicated in the legend. As the convective envelope mass rises above $M_D$, diffusion of $Y$ and $Z$ is turned on in the code and helium is drained from the convective envelope increasing the mass

fraction of hydrogen at the surface. The top axis shows the mass fraction of hydrogen, $X_c$ in the core.

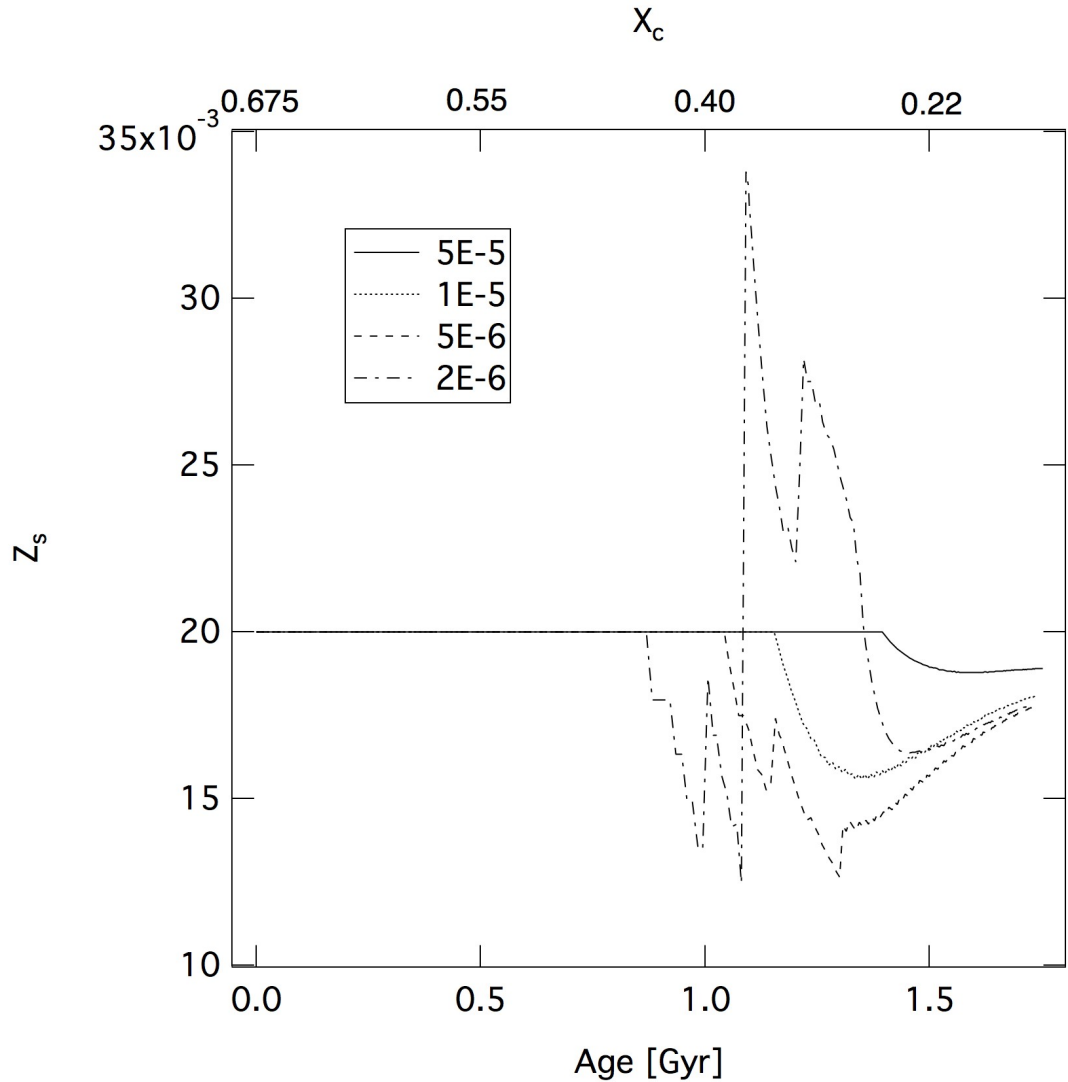

Fig. 3 — The time evolution of the surface mass fraction of metals, $Z_s$ for a 1.5 $M_\odot$ evolutionary track. The different curves correspond to different $M_D$. As the convective envelope mass rises above $M_D$ diffusion of $Y$ and $Z$ is turned on in the code and metals are initially drained from the convective envelope. For $M_D = 2 \times 10^{-6}$ $M_\odot$ the evolution

code cannot stabilize the location of the base of the convective envelope resulting in large fluctuations in the surface abundance of metals. The top axis shows the mass fraction of hydrogen, $X_c$ in the core.

Simply stated, we are not yet able to correctly model, with fully justified physics, diffusion through phases of evolution where the convective envelope mass drops below $\sim 1\times 10^{-6}$ $M_\odot$. Consequently, we chose not to include diffusion in our grids comparing the three different core overshoot models. We have, though, computed a grid of models (raOvD) similar to the raOv grid but with diffusion turned on whenever the mass of the convective envelope rises above $M_D = 2\times 10^{-5}$ $M_\odot$ to see how our results may be affected by the presence of diffusion in the models.

## 4. ASTEROSEISMIC DATA

Two data reductions of the ground based observations have been produced by Bedding et al. (2010), scenario A and scenario B. Although it is easy to identify the $l = 0$ and 1 ridges in the power echelle diagram (Fig. 4, Bedding et al. 2010) it is much more difficult to determine which vertical ridge is $l = 0$ and which is $l = 1$. There are hints of power peaks running along both sides of the two primary ridges that could represent $l = 2$ or higher order modes or could represent bumped modes or aliasing-like effects. Consequently, the ambiguity in assigning the even and odd $l$-value ridges leads to two reductions, labeled scenario A and B. The two scenarios, though, do not yield similar best model fits. Where scenario A yields both self-consistent and reasonable model fits to Procyon, scenario B does not.

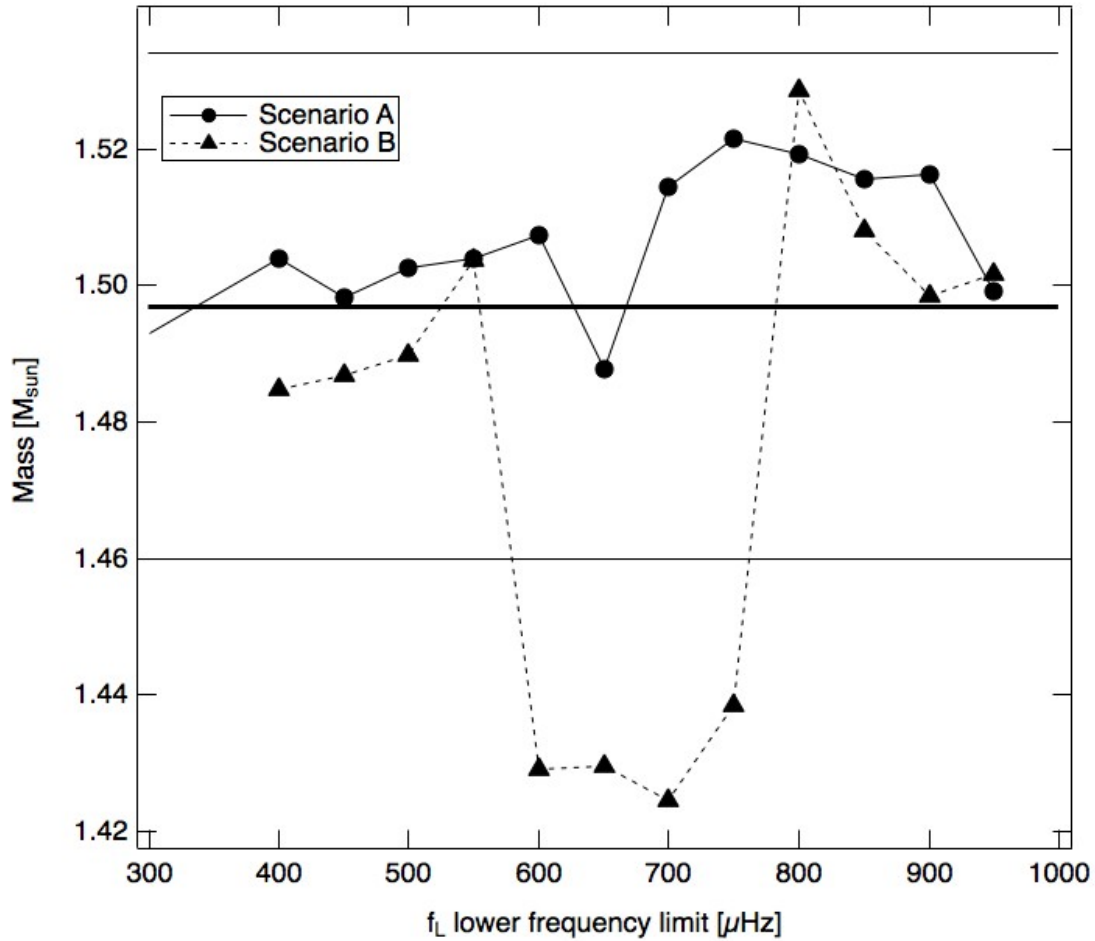

Fig. 4 — The mass of the most probable model when constrained by the asteroseismic data versus the lower frequency limit of the data, $f_L$. Both scenario A and scenario B asteroseismic data reductions (from Bedding et al. 2010) are used. Note that the mass of the most probable model depends on the lower frequency limit of the data when the scenario B asteroseismic data reduction is used to constrain the models. The observed mass of Procyon is indicated by the thick solid horizontal line, with parallel lines showing the uncertainty limits. As the lower frequency limit is extended to include more of the lower frequencies in each data set the mass of the most probable model varies considerably for scenario B but not for scenario A.

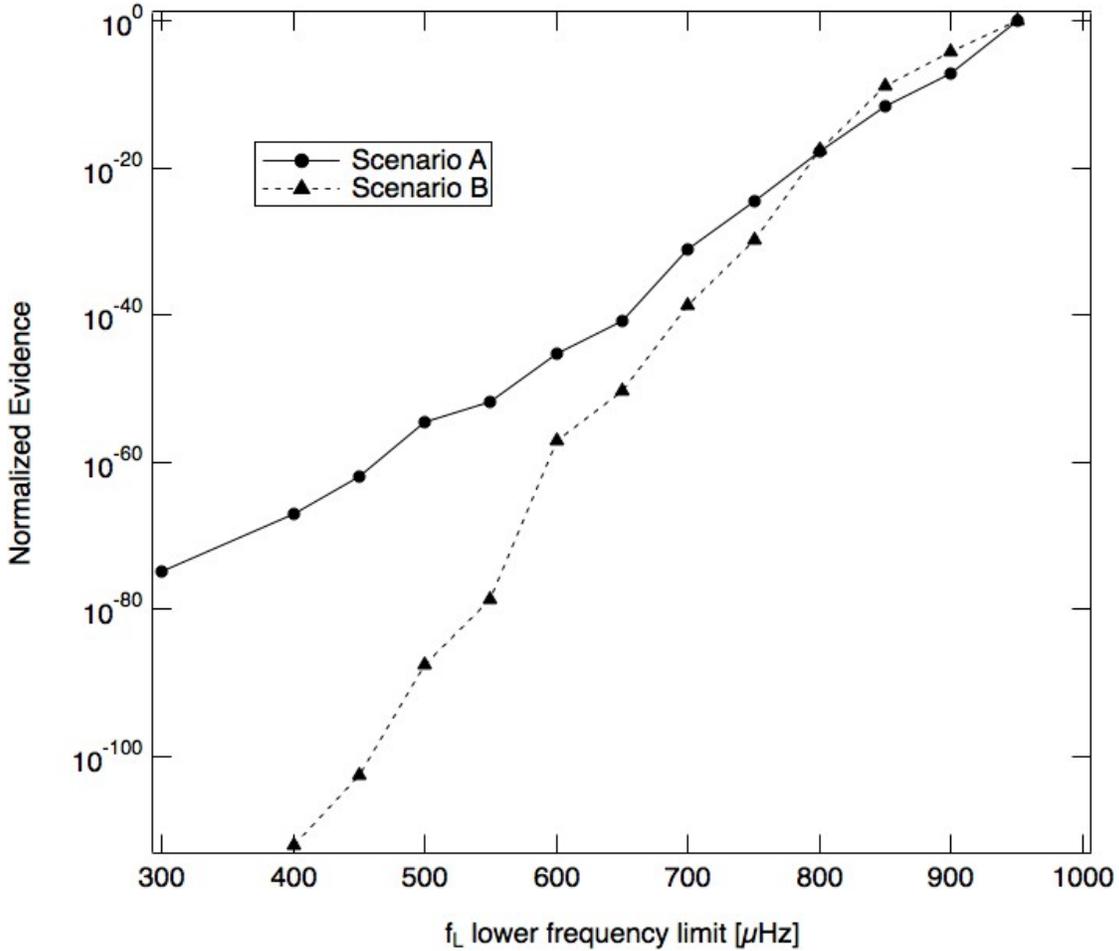

Fig. 5 — The normalized Bayesian evidence versus the lower frequency limit, $f_L$, of the two observed data sets, scenario A and B (see definitions in text). The normalized evidence does not drop off as sharply for scenario A as for scenario B suggesting that at lower frequencies there are fewer inconsistencies between the observations and our models for scenario A than scenario B. The evidence for $f_L = 300\mu$Hz in scenario B is below the lower numerical limit of our Bayesian code and is not shown.

To test the self consistency of each set of frequencies we computed best fit models to several subsets of the frequencies. For simplicity of presentation, here we only show one of the many subset sequences we considered. For our sequence of subsets we selected modes with frequencies above a lower limit frequency $f_L$ for $f_L$ = 300, 400, 450, …900 µHz. The data do not contain frequencies below 300 µHz, so that $f_L$ = 300 µHz corresponds to using the entire data set. The mass of the best fit model matching the observations as a function $f_L$ is plotted in Fig. 4 for scenario A and B. The best fit models were determined as the most probable models using our Bayesian search software. To simplify our argument only the $\beta$ = 0.0, i.e., no overshoot, models in the raOv grid are used. Our conclusions do not change when overshoot and diffusion are included in our comparisons. Fig. 4 reveals the fundamental problem with scenario B, that is, as more of the lower frequencies are included, the mass of the best fit model changes abruptly. We suspect the drop and rise in the mass of the fitted models for scenario B is caused by errant $p$-mode frequencies somewhere in the range 600-750 µHz. As the errant frequency(ies) are included in the model fits the best model's mass is perturbed, until at lower frequencies either correct or oppositely compensating frequencies are included in the model fitting. The mass of the best models determined by different subsets from scenario A remain consistent, all falling near the observed mass = 1.497±0.037 $M_\odot$. We also find that the age, composition, mixing length parameter, and HR diagram location determined from a variety of different subsets of the asteroseismic data vary significantly for scenario B but do not for scenario A.

In Fig. 5 we plot the normalized evidences of the best model fits as a function of $f_L$ for the two scenarios. The evidences are normalized to their peak value at $f_L$ = 950 µHz. Note that for scenario B at the lowest value of $f_L$, the evidence falls below the lower numerical limit of the code. As more lower frequencies are included in the model fitting the evidences drop indicating that either the lower frequencies are less well determined than the higher frequencies or that our models are not completely accurate in the deeper regions of the model where the lower frequencies penetrate. Regardless, the normalized evidences for scenario B decrease significantly more rapidly than scenario A. This again suggests that some of the lower frequencies in scenario B are incorrectly identified.

We carried out extensive model testing using both scenario A and B. Scenario A produced self-consistent and reasonable results and scenario B produced self-inconsistent and confusing results that were also inconsistent with other observable constraints. If Scenario B is, in fact, correct, then it is well outside our ability to produce models that fit it. To fit models to Scenario B will require a significant change to our standard assumptions of stellar evolution for Procyon. Hereafter, we use Scenario A for our modeling analyses.

## 5. SENSITIVITY TO MODEL PARAMETERS

To test the sensitivity of our models to some of the key physical parameters, we computed a variety of stellar models for Procyon differing from each other by a single perturbed parameter. All of the models were evolved from the ZAMS to the observed position of Procyon in the HR diagram.

For each evolutionary sequence, the mass fraction of helium, $Y$, was adjusted, i.e., tuned, so that the evolutionary track passed through Procyon's HRD location. An evolutionary track is computed with a trial $Y$ and the track's offset from Procyon's $T_{eff}$ and $L$ is stored. Another track is then computed with a slightly different $Y$ and that track's offset from Procyon's $T_{eff}$ and $L$ is combined with the stored value from the previous track to compute, via linear interpolation, a new value for $Y$ that will zero the offset, that is, produce a track that passes directly through Procyon's HR diagram location. The code will repeat the process, iterating until the track passes through Procyon's $T_{eff}$ and $L$ within the uncertainties. The model along this track that falls closest to the $T_{eff}$ and $L$ values for Procyon is then selected and its pulsation spectrum computed.

For reference we define a standard model having a mixing length parameter $\alpha = 2.0$, a mass equal to 1.497 $M_\odot$, $Z = 0.02$, no diffusion, and no overshoot. Note that the reference model is not to be interpreted as the best model fit to Procyon using standard physics. It simply represents the model we get from a basis set of model parameters. Indeed, we note that the Z/X for this model is too high compared to the observations. One parameter of the standard model was then changed to see its effect on the *p*-mode oscillation frequencies and on the structural properties of the model itself. The *p*-mode frequencies were not used to constrain the models, only the HRD location and mass.

In Table 3 we list some of the fundamental properties of the models. The standard or reference model heads the list. All the models fit Procyon's location in the HR diagram and have mass = 1.497 $M_\odot$, initial metal mass fraction $Z_0 = 0.02$, and mixing length parameter $\alpha = 2.0$, The rest of the properties of the models, listed left to right, are: $\beta$, the amount of core overshoot in pressure scale heights; $M_D$, the minimum threshold mass, for diffusion (see discussion section 3); $Y_0$, the initial homogeneous abundance of helium on the ZAMS; $Z_s/X_s$, the surface mass fraction ratio of metals to hydrogen; $Z_s$, the surface mass fraction of metals; age, the age of the model in Gyr; $M_{cc}$, the mass of the convective core in units of $M_\odot$; $x_{ce}$, the radius fraction location of the base of the convective envelope; $X_c$, the central hydrogen mass fraction, log $P_c$, the log of the central pressure in dyne cm$^2$, log $T_c$, the log of the central temperature in K, and log $\rho_c$, the log of the central density in g cm$^2$. The *env* model has convective envelope overshoot below the base of the convective envelope of 0.5 pressure scale heights.

TABLE 3

MODEL PARAMETERS - DIFFERENTIAL STUDY

| ID | β | $M_D$ | $Y_0$ | $Z_s/X_s$ | $Z_s$ | age | $M_{cc}$ | $X_{ce}$ | $X_c$ | $\log P_c$ | $\log T_c$ | $\log \rho_c$ |
|---|---|---|---|---|---|---|---|---|---|---|---|---|
| *ref* | 0.0 | - | 0.302 | 0.0295 | 0.0200 | 1.787 | 0.107 | 0.865 | 0.076 | 17.358 | 7.366 | 2.146 |
| *diff1* | 0.0 | 5E-5 | 0.302 | 0.0270 | 0.0189 | 1.753 | 0.110 | 0.867 | 0.092 | 17.346 | 7.361 | 2.129 |
| *diff2* | 0.0 | 1E-5 | 0.302 | 0.0252 | 0.0181 | 1.736 | 0.111 | 0.869 | 0.109 | 17.337 | 7.357 | 2.114 |
| *diff3* | 0.0 | 5E-6 | 0.302 | 0.0243 | 0.0177 | 1.730 | 0.115 | 0.870 | 0.116 | 17.334 | 7.355 | 2.108 |
| *diff4* | 0.0 | 2E-6 | 0.302 | 0.0244 | 0.0178 | 1.721 | 0.114 | 0.870 | 0.120 | 17.333 | 7.354 | 2.105 |
| *ov1* | 0.3 | - | 0.287 | 0.0289 | 0.0200 | 2.066 | 0.135 | 0.867 | 0.220 | 17.311 | 7.338 | 2.043 |
| *ov2* | 0.6 | - | 0.270 | 0.0282 | 0.0200 | 2.449 | 0.157 | 0.869 | 0.313 | 17.304 | 7.327 | 2.002 |
| *ov3* | 0.9 | - | 0.255 | 0.0276 | 0.0200 | 2.869 | 0.172 | 0.872 | 0.380 | 17.302 | 7.320 | 1.976 |
| *ov4* | 1.2 | - | 0.241 | 0.0271 | 0.0200 | 3.311 | 0.183 | 0.873 | 0.425 | 17.301 | 7.316 | 1.960 |
| *ov5* | 1.5 | - | 0.227 | 0.0266 | 0.0200 | 3.807 | 0.191 | 0.875 | 0.461 | 17.301 | 7.313 | 1.948 |
| *env* | 0.0 | - | 0.278 | 0.0289 | 0.0200 | 2.179 | 0.073 | 0.868 | 0.008 | 17.541 | 7.412 | 2.328 |

The effect that parameter changes have on the radial ($l = 0$) *p*-mode frequencies is shown in figures 6 and 7. Fig. 6 plots the perturbation in frequency of the diffusion models (*diff1, diff2, diff3, diff4*) and the convective envelope overshoot model (*env*) with respect to the nonadiabatic frequencies of the *ref* model as a function of *n* or frequency. Fig. 7 plots the perturbation in frequency of the core overshoot models (*ov1, ov2, ov3, and ov4*). Higher frequency modes are more sensitive to the surface layers because they, themselves, are confined to the outer envelope, with shallow inner turning points. Lower frequencies have deeper inner turning points, hence, are also affected by the structure at deeper depths. To help provide a sense of the physical scale for the x-axis in Figs. 6 and 7 we include a plot of the adiabatic minus the nonadiabatic frequencies for the standard model. For higher frequencies the adiabatic frequencies are lower than the nonadiabatic frequencies, as is the case for the solar model but the magnitude of the adiabatic minus nonadiabatic perturbation is one-quarter that for the solar model (the frequency shift at the higher frequencies for a solar model is of the order 10 to 20 µHz, Guenther 1994).

Diffusion of $Y$ and $Z$, as helium and metals are drained out of the surface convective envelope, primarily affect the surface layers by decreasing the mean molecular weight in the region and by perturbing the opacities. Diffusion, in addition to decreasing the surface abundances of metals and helium, decreases the age by ~3%, and increases the mass of the convective core between 3% and 8% depending on $M_D$. As different values of $M_D$ do affect the model, it will be important in the future to properly constrain the rate of diffusion in models that have thin convective envelopes. Diffusion does not significantly affect the position of the base of the convective envelope.

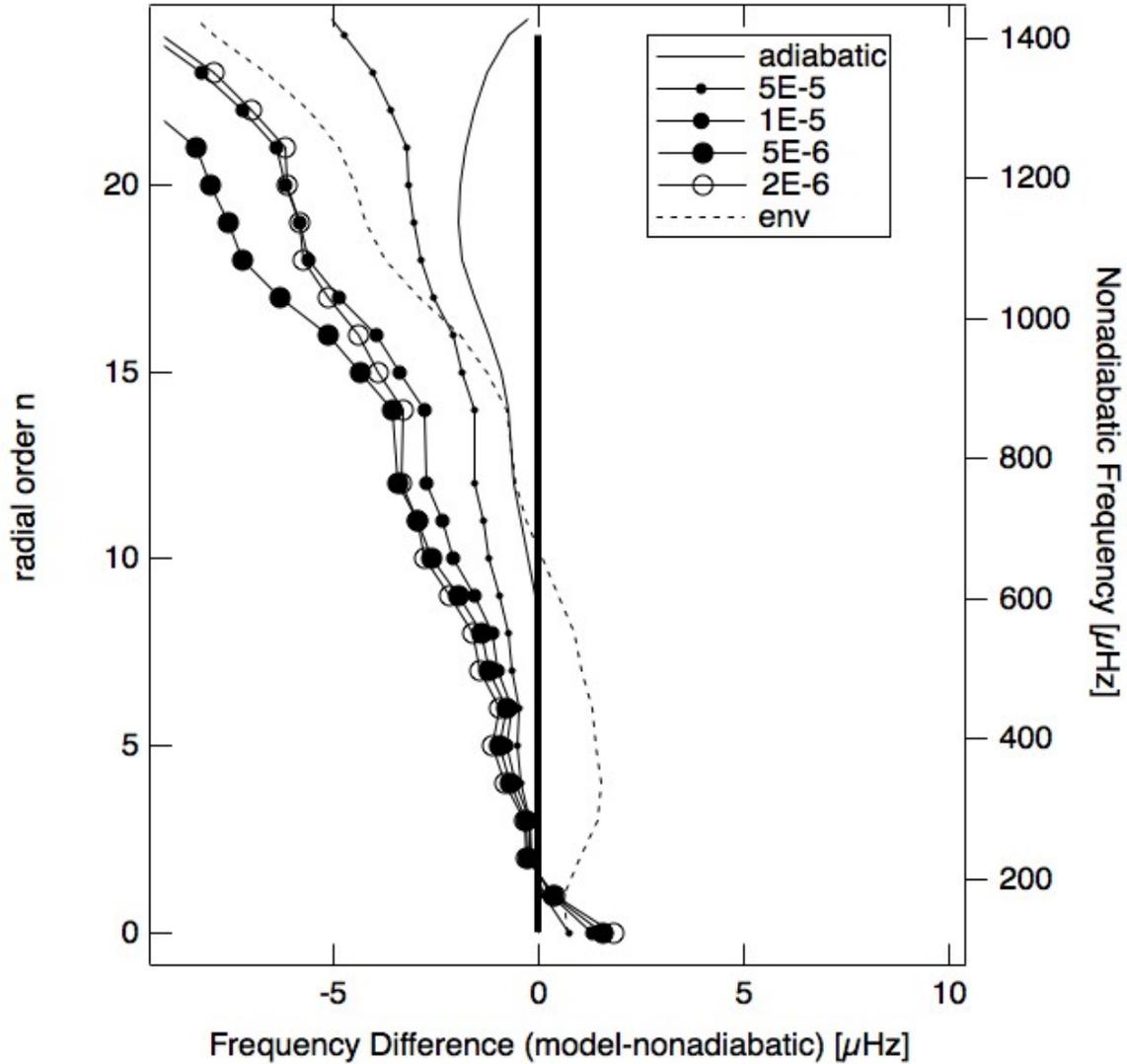

Fig. 6 — Shown for radial modes, the nonadiabatic frequency difference between tuned models of Procyon that include diffusion and a reference model (*ref*) that does not include diffusion are plotted against the radial order *n*, with the corresponding nonadiabatic frequency shown on the right. The frequency difference between a model with 0.5 pressure scale height convective envelope overshoot (*env*) is also plotted along with the adiabatic minus nonadiabatic frequencies of the reference model. The different diffusion models with different values of convective envelope mass cutoffs ($M_D$) are indicated in the legend, with increasing plot symbol size corresponding to decreasing $M_D$.

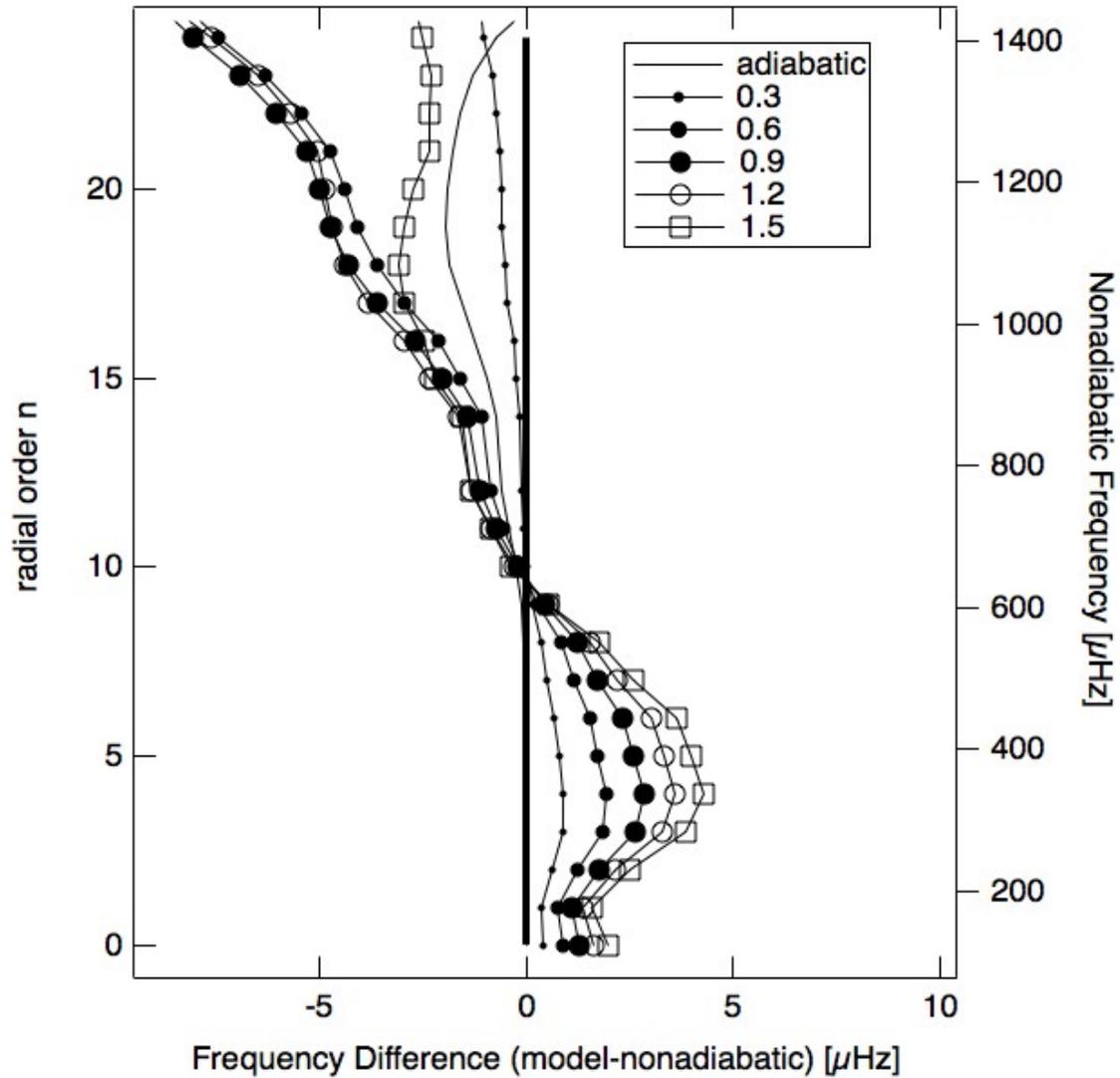

Fig. 7 — Shown for radial modes, the frequency difference between tuned models of Procyon that include convective core overshoot and a reference model (*ref*) that does not include core overshoot are plotted against the radial order *n*, with the corresponding nonadiabatic frequency shown on the right. The adiabatic minus nonadiabatic frequencies of the reference model is also plotted. The extent of core overshoot β is indicated in the legend with increasing plot symbol size, then open circle and open square corresponding to increasing amounts of overshoot.

Not too surprisingly, diffusion does perturb the *p*-mode frequencies. The frequencies of the modes confined to the outer layers decrease, a consequence of the lowered mean molecular weight. Diffusion has little effect on the lower frequencies.

Core overshoot, β, increases the mass of the convective core, with the effect itself increasing with increasing β, i.e., the extent of core overshoot. The age is significantly affected by the presence of core overshoot. The larger the convective core the more hydrogen is mixed into the core lowering the mean molecular weight in the region, which in turn lowers the rate of nuclear burning and, indirectly, the luminosity of the model at a given phase of evolution. As a consequence, the stellar model takes longer to reach the observed luminosity and effective temperature location for Procyon. Although our models of Procyon have not yet reached core hydrogen exhaustion, seen as turn-off in the HR diagram, the increased supply of hydrogen to the core will ultimately delay turn-off. Because age determinations of stars in clusters depend on the location of turn-off, the extent of core overshoot used in stellar model isochrones will affect the determined age. Note that as the overshoot amount increases (between *ov4* and *ov5*), the initial abundance of helium begins to drop below the primordial abundance.

In order to conserve mass and to maintain hydrostatic equilibrium, a decrease in density near the core has to be compensated by an increase in the density in the envelope. In other words, because the models have the same mass and are all tuned to a specific location in the HRD, hence, have identical radii, any perturbation to the run of density in one part of the star must be balanced by the opposite change in the rest of the star. For the lowest frequency modes that extend deep into the interior, the lower densities in the core shift the frequencies of each mode to higher values (frequency is inversely proportional to the square root of the density). Shallower penetrating modes only see the more dense, compared to the standard model, outer envelope regions, hence, are perturbed in the opposite sense. The effect on the higher frequencies of every increasing amounts of core overshoot is nonlinear with the relative change in perturbation decreasing. This is because the values of $Z$ and $\alpha$ are fixed in these comparisons and the large values in overshoot can only be compensated by equivalently large changes in $Y$. But changes in $Y$ affect the opacities and perturb the location of the helium ionization zone, which in turn perturbs the frequencies.

The shape of the perturbation curves for diffusion and for core overshoot are distinct, hence, it should be possible to constrain both values independently using accurately observed $p$-mode frequencies.

Fig. 6 also shows the effect of convective envelope overshoot of 0.5 pressure scale heights (*env*). The *env* curve is similar in shape to the core overshoot curve. The models, themselves, though, are, in fact quite distinct. Table 3 shows that the *env* model is very near hydrogen core exhaustion. It will produce distinct frequency difference curves for higher $l$-value modes. Here one can think of this effect on the higher $l$-value frequencies as due to differences in the small frequency spacing between the *ov* and *env* models. We, therefore, expect that when we are able to extend our grids to include additionally a range of envelope overshoot values that we will be able to uniquely constrain not only the core overshoot but also the envelope overshoot amount. Using our existing computational resources and software tools, it would take a year to compute these grids.

## 6. MOST PROBABLE MODEL FITS TO PROCYON

We computed four distinct model grids, each spanning a wide range of mixing length parameter, composition, age, log $L/L_\odot$, log $T_{\text{eff}}$, mass, and core overshoot amount. Three of the grids, as described in section 2.3, test three different convective core overshoot formulations, raOv: standard MLT and core overshoot with a radiative temperature gradient; adOv: standard MLT and core overshoot with an adiabatic temperature gradient; and Ku: nonlocal Kuhfuss MLT (core overshoot has an adiabatic temperature gradient). The raOvD grid is similar to the raOv grid except that $Y$ and $Z$ diffusion is turned on whenever the convective envelope mass is greater than $M_D = 2 \times 10^{-5}$ $M_\odot$, which occurs ~1Gyr after the ZAMS (see Fig. 1).

In general we compare how well the models in each subgrid (a function of β, the amount of core overshoot) fit the observations using Bayesian probabilities. We compute and compare the evidence for each subgrid, identifying the most probable model within each subgrid. Our most probable models will deviate slightly from a chi-squared minimization determination of the best fit model because the Bayesian approach incorporates, in a probabilistic self-consistent manner, the HR diagram location and mass of Procyon and the possible existence of surface effects. The Bayesian evidence quantifies the comparison between most probable model fits within each subgrid, with evidence ratios greater than 10 deemed significant.

In figure 8 we compare model fits based on the adiabatic *p*-modes to model fits based on the nonadiabatic *p*-modes for the raOv grid and we compare the model fits that include the mass and HR diagram priors to those that do not, again for the raOv grid. The evidence for the model fits using nonadiabatic *p*-modes are significantly greater than the model fits using adiabatic *p*-modes. The odds ratios, nonadiatic/adiabatic, vary from 10 to 100 for each overshoot subgrid. Note that the nonadiabatic *p*-mode calculation accounts for radiative gains and losses only. We found similar evidence ratios favoring the nonadiabatic *p*-modes for the other grids, Ku, adOv, and raOvD.

Fig. 8 also shows that the evidences are not significantly affected by the HR diagram and mass priors. This implies that the asteroseismic data alone are yielding model fits to the data that, independently, fit the HR diagram and mass constraints for Procyon.

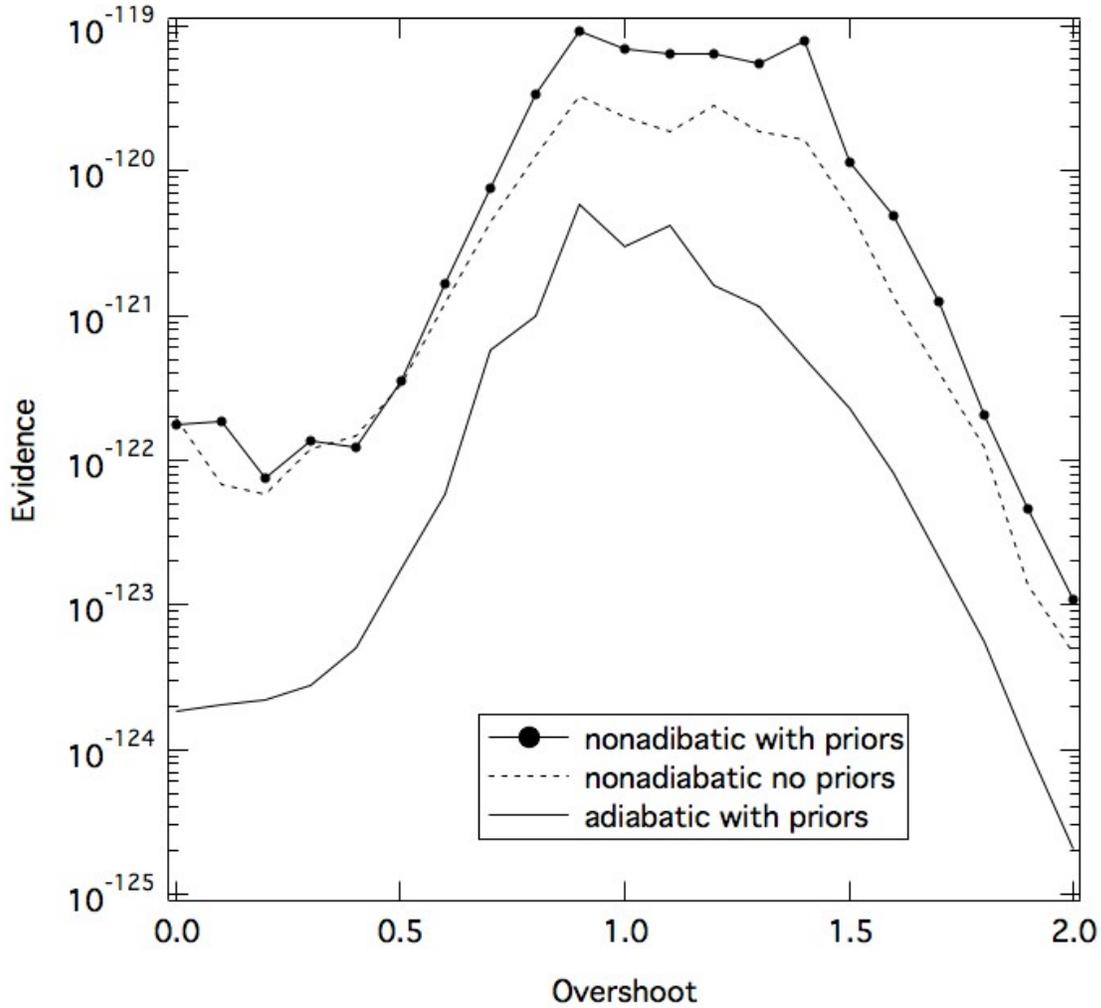

Fig. 8 — Evidence versus core overshoot, $\beta$, for models from the raOv grid. The legend identifies whether the models fits used adiabatic *p*-modes or nonadiabatic *p*-modes and whether the HR diagram and mass priors were used or not. Evidence ratios greater than 10 are considered significant.

In figure 9 we compare the evidences of the model subgrids as a function of overshoot for the raOv, raOvD, adOv, and Ku grids. To save computational time the adOv and Ku grids were not extended to overshoot values beyond $\beta = 1.0$. The evidences initially increase for all models of overshoot. The evidence for Kuhfuss overshoot levels off at $\beta = 0.5$. The evidence for the standard MLT overshoot with the overshoot regions forced to be adiabatic (adOv) peaks at $\beta = 0.9$ then abruptly drops off. Regardless, both the adOv and Ku model grids have significantly lower evidences than the raOv and raOvD models. Both the standard MLT overshoot models (raOv) and the standard MLT overshoot models with $Y$ and $Z$ diffusion (raOvD) show significantly higher values of evidence for $\beta$ between 1.0 and 1.5 than any other model subgrid. Further, the models with diffusion (raOvD) are systematically preferred according to the evidence over those that do not.

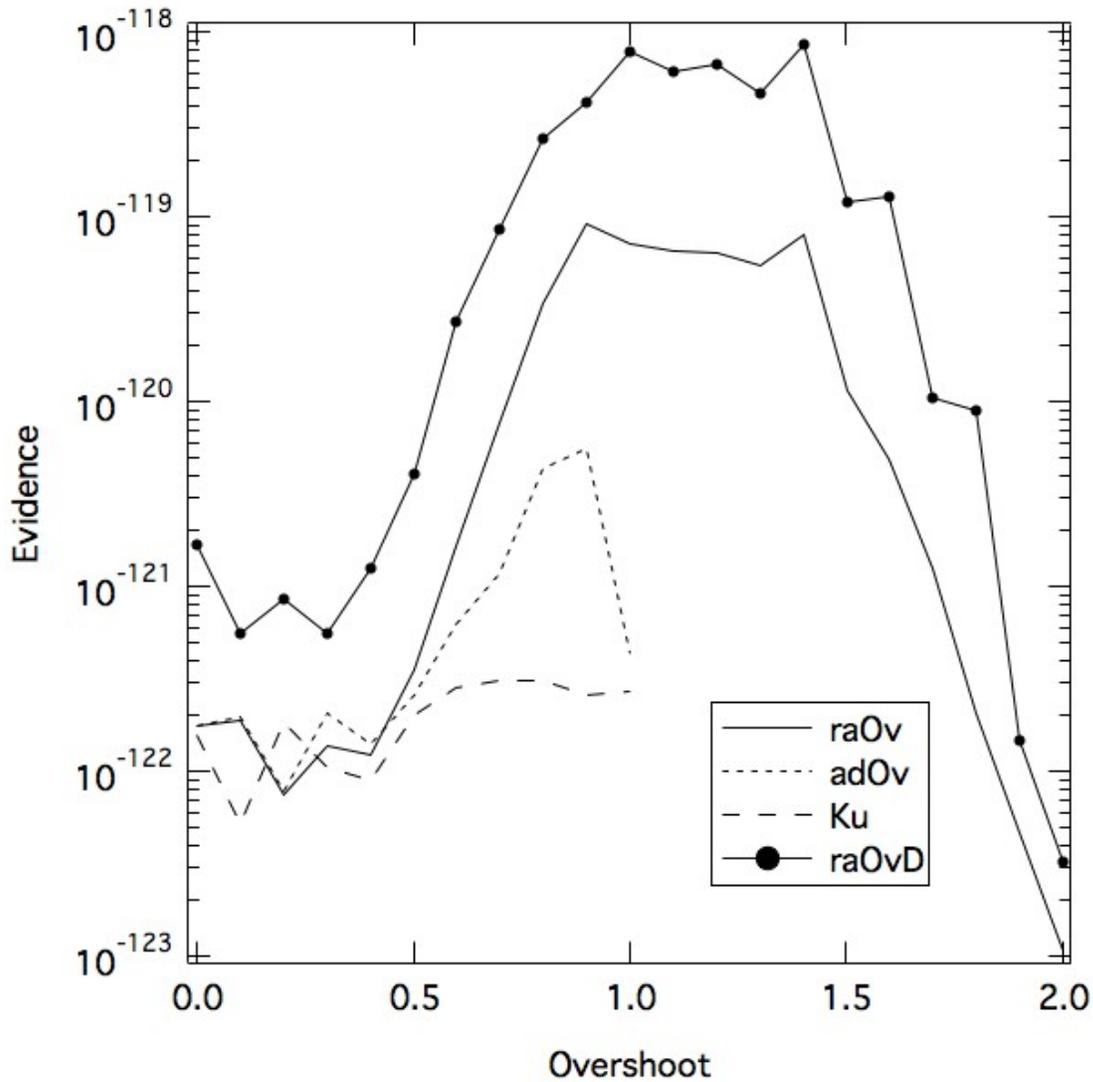

Fig. 9 — The evidence versus core overshoot, β, for the raOv, standard MLT overshoot, adOv, standard MLT overshoot with the overshoot region forced to be adiabatic, Ku, Kuhfuss nonlocal MLT, and raOvD, standard MLT overshoot but with diffusion of helium and metals also included in the model calculation. Note that the Ku and adOv grids do not extend beyond β = 1.0.

The evidences strongly support the following model based conclusions for Procyon:

1. Although using Kuhfuss overshoot does slightly improve the models for low values of overshoot, compared to the other overshoot models, Kuhfuss falls significantly behind. For the case of Procyon and the physics here considered, the extra computational effort to compute Kuhfuss overshoot is not justified.

2. Models with a forced adiabatic overshoot region do not fit Procyon as well as models that maintain a radiative gradient in the overshoot region, implying that the core overshoot region is unlikely to be adiabatic. This is also supported by the lower evidence for the Kuhfuss model which also has an adiabatic overshoot region. This conclusion agrees with the findings, noted in the introduction, from detailed 3D numerical calculations of convection in stars.

3. Based on the raOv and raOvD evidence curves, the asteroseismic data strongly supports the existence of convective core overshoot in Procyon, with β between 1.0 and 1.5 pressure scale heights. This suggests that the asteroseismic data on other stars of equivalent or greater quality will also be able to test for the existence of core overshoot.

4. Because the raOv and raOvD evidence curves run parallel to each other both peaking in the same range of overshoot, we confirm that the Bayesian search software is correctly handling the surface effects and not allowing the surface effects to skew the overall asteroseismic fits to the data.

5. Because diffusion does improve the models, we conclude that there are modeling physics near the surface, such as diffusion, that need to be included to provide the best model fits to Procyon. We are reluctant to conclude at this time that the effects are solely or even partially due to diffusion because of the known weaknesses in our model of diffusion and because there also exists the possibility that convective envelope overshoot exists, which could mask or mimic the effects of diffusion.

The mass of the most probable model in each subgrid, as shown in Fig. 10, is within one sigma of the observed mass. The HR diagram positions of the most probable model from each subgrid, see Fig. 11, also for the most part lie within one sigma of the observed HR diagram. Several points in the raOvD grid lie outside the one sigma error box but have high values of overshoot and lower overall evidences. Overall, though, all the models that fit well the asteroseismic data also fit within uncertainties the other observables, supporting our earlier stated impression of the consistency of the asteroseismic data.

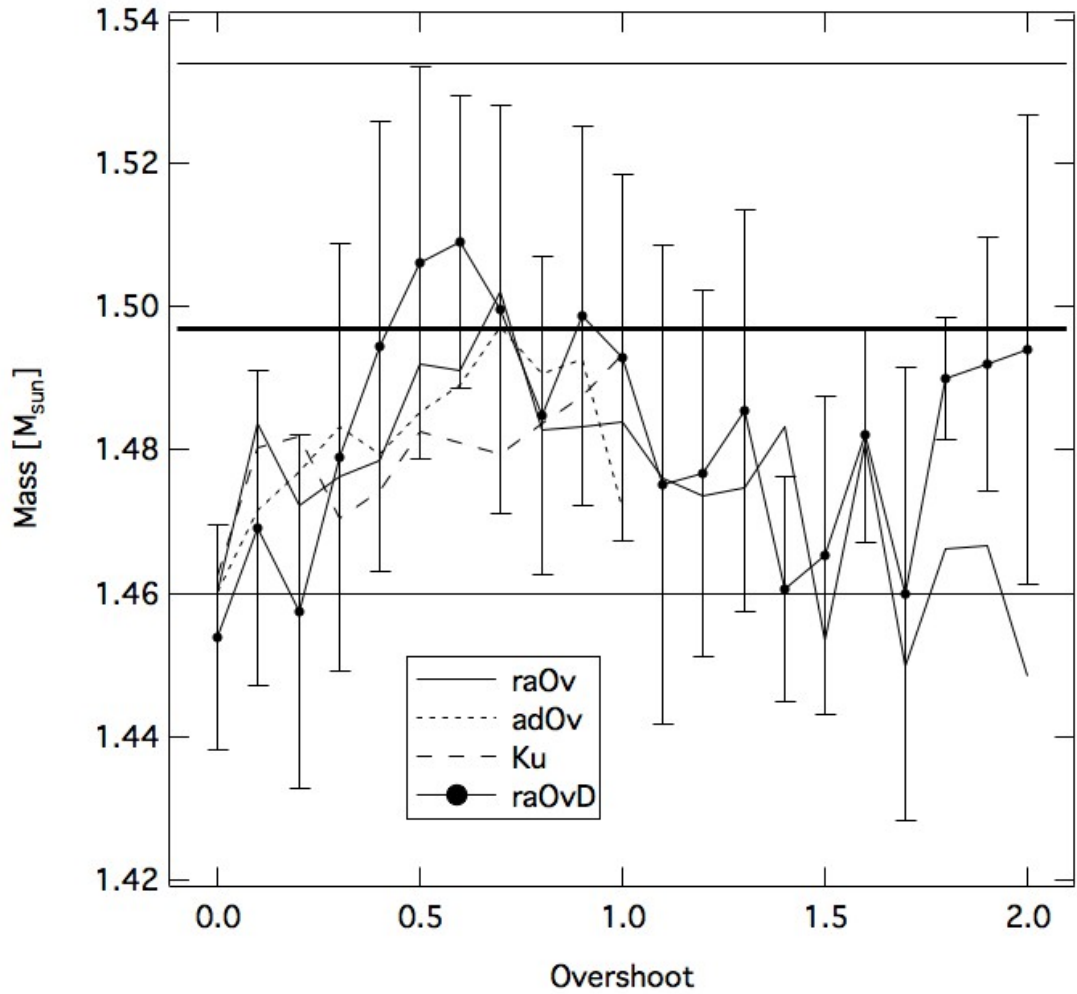

Fig. 10 — Mass of the most probable model in each of the raOv, adOv, Ku, and raOvD subgrids versus core overshoot, β. Error bars are only attached to the raOvD data points but are representative of the uncertainties for the other subgrids. The observed mass (thick line) and one sigma uncertainty range are indicated by the horizontal lines.

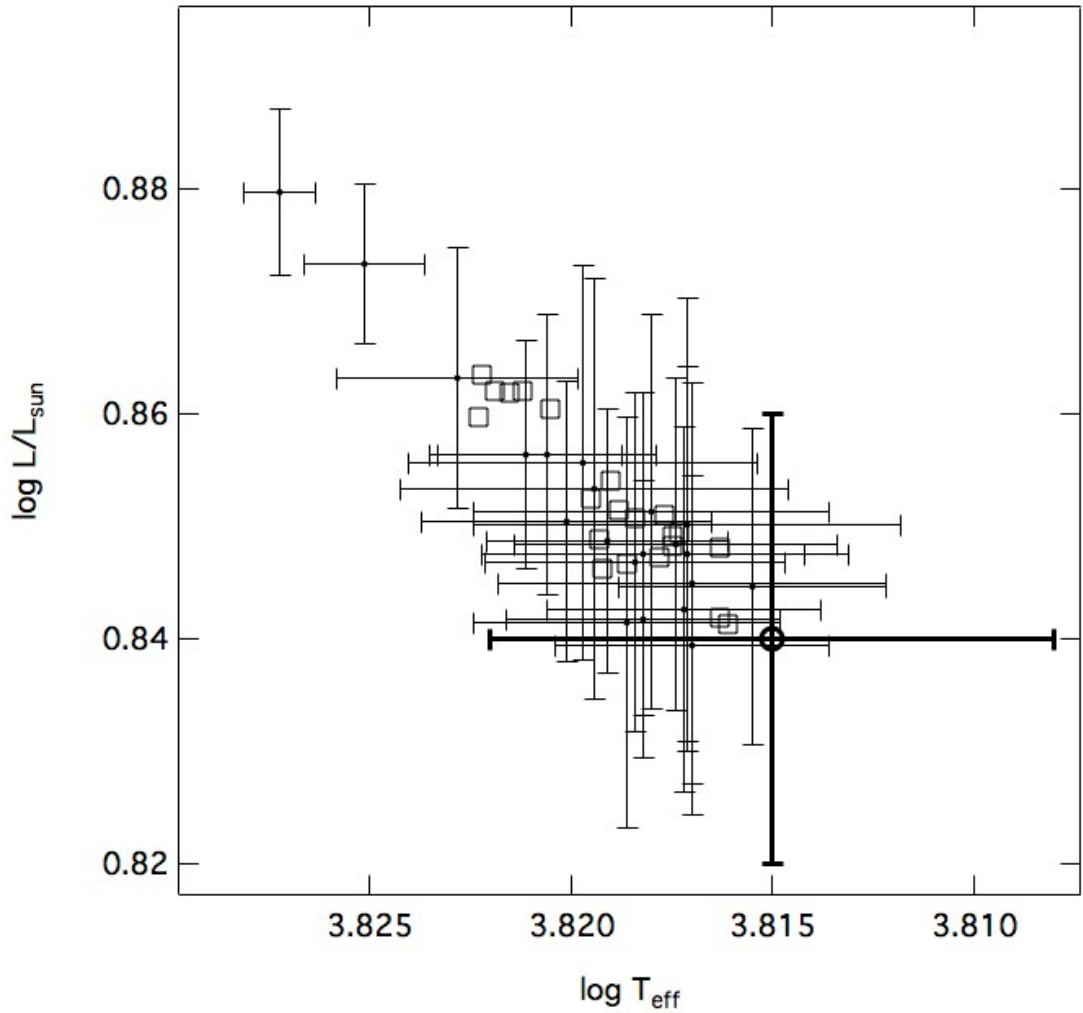

Fig. 11 — HR diagram showing location of the most probable models from the raOvD subgrids (points with error bars) and the raOv (open squares). The three raOvD points falling outside the one sigma error bars marking Procyon's observed position correspond, from far left inward, to overshoot $\beta$ = 2.0, 1.9, and 1.6.

As described in section 5, increasing the amount of overshoot does affect the mass of the convective core and the age. We can also see this behavior in the most probable models. Fig. 12 shows the mass of the convective core increasing with the amount of overshoot and Fig. 13 shows the age of the model also increasing with the amount of overshoot. Since the most probable models overall are the raOvD models with overshoot ranging from $1.0 \leq \beta \leq 1.5$, we conclude that the most probable range of convective core mass for Procyon is 0.18 $M_\odot$ to 0.19 $M_\odot$ and the most probable range of age is 2.4 Gyr to 2.8 Gyr.

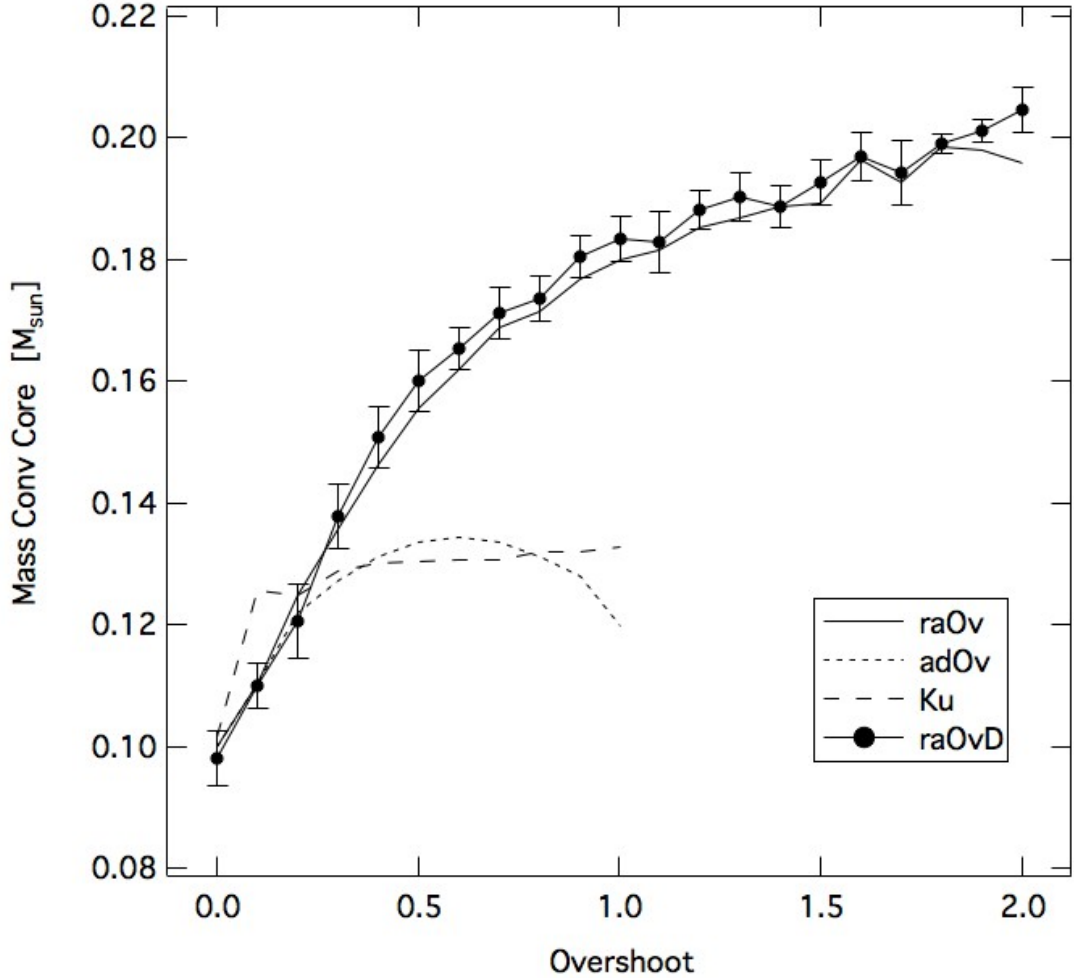

Fig. 12 — Mass of the convective core of the most probable model in each of the raOv, adOv, Ku, and raOvD subgrids versus core overshoot, $\beta$.

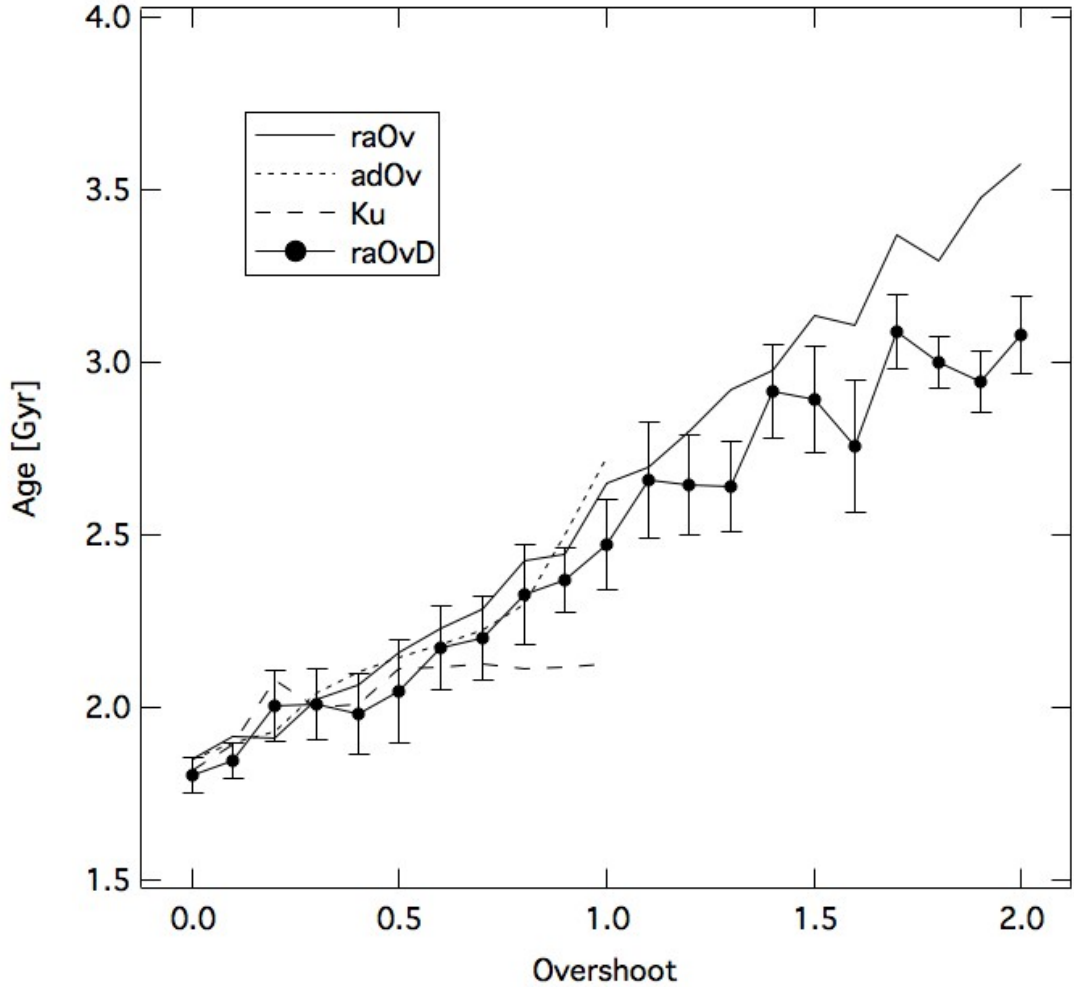

Fig. 13 — Age of the most probable model in each of the raOv, adOv, Ku, and raOvD subgrids versus core overshoot, $\beta$.

We find from the most probable models that the radius fraction of the base of the convective envelope for Procyon is ~0.92±0.01, a result that is relatively insensitive to the choice of overshoot formulation and amount. We also find that the mixing length parameter, also relatively insensitive to the choice of overshoot formulation and amount, is 1.8±0.1. For reference, our standard solar model based on the same physics used to construct our grids has a mixing length parameter that is remarkably close at 1.7.

The surface and initial mass fraction ratio of metals to hydrogen are shown in Fig. 14 (The surface and initial values are, of course, identical for the non-diffusion models). At higher overshoot values for the raOv and adOv models, the surface $Z_s/X_s$ is greater than the observed value. When diffusion is included (raOvD models) and for core overshoot between 0.3 and 1.7, the surface abundances fall back within the uncertainties.

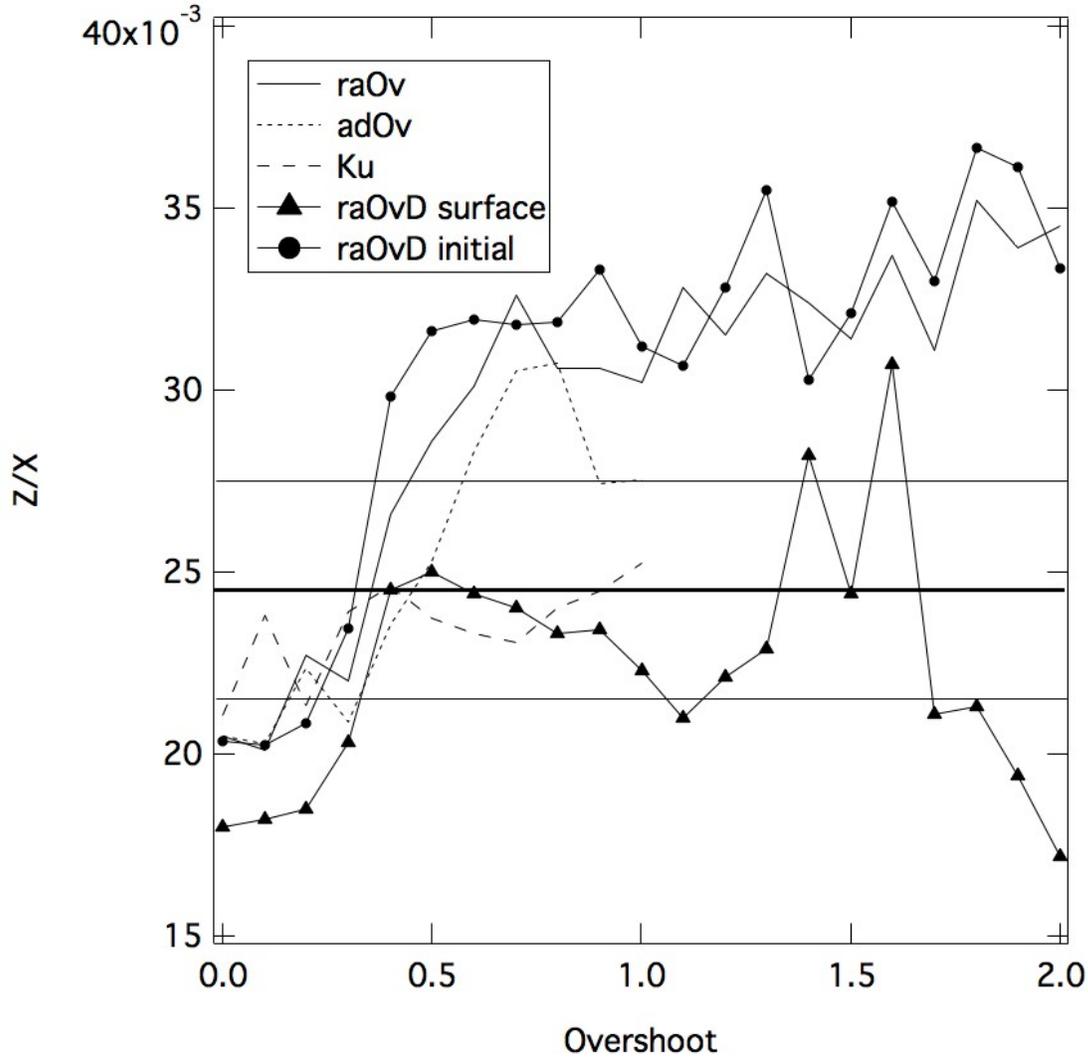

Fig. 14 — The initial and surface mass fraction ratio $Z/X$ for the most probable models in each subgrid versus core overshoot $\beta$. The observed $Z/X$ is indicated along with uncertainty range by the horizontal parallel lines. The surface and initial values of $Z/X$ are the same for the non-diffusion models.

TABLE 4

Parameters of Most Probable Models

| Parameter | raOv β=0.0 | raOv β=1.0 | raOvD β=1.0 |
|---|---|---|---|
| $Y$ & $Z$ Diffusion | No | No | Yes |
| Overshoot [$H_P$] | 0.0 | 1.0 | 1.0 |
| Mass [$M_\odot$] | 1.46±0.03 | 1.48±0.03 | 1.49±0.03 |
| log $L/L_\odot$ | 0.85±0.02 | 0.85±0.02 | 0.86±0.02 |
| log $T_{eff}$ | 3.819±0.04 | 3.817±0.05 | 3.820±0.04 |
| log $R/R_\odot$ | 0.310±0.003 | 0.313±0.003 | 0.312±0.002 |
| $Z_s/X_s$ | 0.020±0.002 | 0.030±0.005 | 0.022±0.002 |
| Age [Gyr] | 1.85±0.08 | 2.65±0.15 | 2.47±0.13 |
| $Y_0$ | 0.282±0.014 | 0.269±0.018 | 0.278±0.015 |
| $Z_0$ | 0.014±0.001 | 0.021±0.003 | 0.022±0.002 |
| $\alpha$ | 1.8±0.1 | 1.8±0.1 | 1.8±0.1 |
| $x_{ce}$ | 0.913±0.006 | 0.910±0.008 | 0.917±0.004 |
| $M_{cc}/M$ | 0.069±0.003 | 0.122±0.008 | 0.123±0.008 |
| $R_{cc}/R$ | 0.059±0.003 | 0.072±0.005 | 0.073±0.005 |
| Evidence | $1.75 \times 10^{-122}$ | $7.06 \times 10^{-120}$ | $7.72 \times 10^{-119}$ |
| $\langle \Delta\nu_0 \rangle$ [μHz] | 55.1±1.0 | 54.8±1.0 | 54.7±0.9 |
| $\langle \Delta\nu_1 \rangle$ [μHz] | 55.0±1.0 | 55.1±1.1 | 55.1±0.9 |
| $\langle \Delta\nu_2 \rangle$ [μHz] | 55.2±1.1 | 55.0±1.0 | 54.9±0.9 |
| $\langle \delta\nu_0 \rangle$ [μHz] | 4.4±0.3 | 3.5±0.5 | 3.7±0.6 |

In table 4 we list some of the parameters of three of our most probable models, i.e., the models that fit the asteroseismic data, HR diagram location and mass of Procyon. The first column corresponds to the best fitting model from the raOv grid that does not include overshoot and does not include $Y$ and $Z$ diffusion. The second column corresponds to a best fitting model from the raOv grid that does include overshoot but does not include diffusion. This model is representative of the range of raOv models from $\beta = 1.0$ to 1.5, all of which have high evidences. And the third column of parameters corresponds to a best fitting model from the raOvD grid that includes both diffusion and overshoot. The fundamental parameters, mass, luminosity, effective temperature, and radius are in general agreement with observations (compare Table 4 to Table 1). The significance of the differences in Z/X are discussed in the conclusions. The mass, the mixing length parameter ($\alpha$), and the depth of the convective envelope ($x_{ce}$) are, within uncertainties, unaffected by the inclusion of diffusion or overshoot. The age and convective core mass, $M_{cc}$, increase, as expected, when overshoot is included. The radius fraction location of the outer edge of the convective core, $R_{cc}/R$, increases by ~22% with the inclusion of overshoot $\beta = 1.0$. The initial metal abundance also increases when overshoot is included, possibly compensating for the effect of the increased core mass on the nuclear burning rates. When diffusion is also included, the best fitting models show depletion of $Z$ (and $Y$) at the surface. The inclusion of diffusion, though, does not significantly alter the convective core mass or age (comparing to models with overshot). The initial abundance of helium $Y_0$ is within uncertainties of the solar value ($Y_0 = 0.2714$) and in agreement with Galactic and cosmological nucleosynthesis predictions.

Finally, to show how well the frequencies of the models do fit the observed frequencies we present, in an echelle diagram for Procyon, the model frequencies for the most probable model without overshoot and diffusion (raOv $\beta = 0.0$) and with (raOvD $\beta = 1.0$).

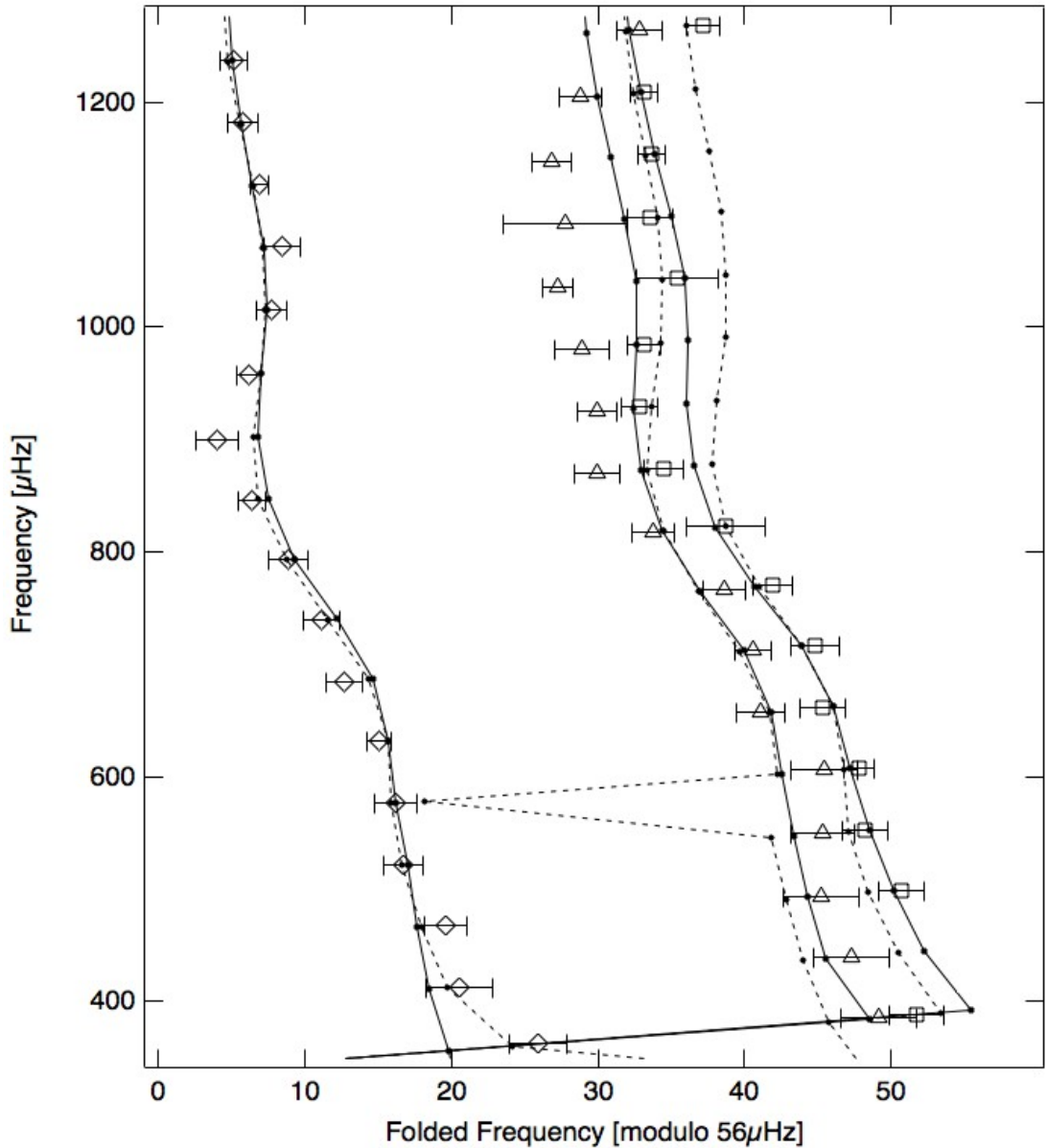

Fig. 15 — Echelle diagram (frequency versus frequency modulo 56μHz) for the scenario A Procyon observations from Bedding et al. (2010) and two examples of most probable model fits to both the observed *p*-mode frequencies and Procyon's observed position in the HR diagram and mass. The $l = 0$ (open squares), $l = 1$ (open diamonds), and $l = 2$ (open triangles) *p*-modes for Procyon are shown with indicated uncertainties. The *p*-mode frequencies of the most probable model of Procyon without diffusion or overshoot is shown with dashed lines and the most probable model of Procyon that includes diffusion and overshoot is indicated with solid lines.

Both models fit the observed $p$-modes well and, as previously noted, also match well Procyon's HR diagram location and mass. But the raOvD $\beta = 1.0$ model does fit the lower and higher frequencies slightly better than the raOv $\beta=0.0$ model. As described in section 5, overshoot affects the lower frequencies and, in the opposite direction, the upper frequencies while diffusion only perturbs the upper frequencies. Including both overshoot and diffusion in the models yields a fit that, according to the Bayesian evidence, is significantly more probable than models that do not have either.

The observed $l = 2$ $p$-mode frequencies above 800 µHz are not fit as well as the $l = 0$ and $l = 1$ curves. Equivalently, we note that the small spacing for the best fitting models with overshoot (raOv and raOvD) are smaller than the observed small spacings (compare the small spacings listed in table 4 to the observed small spacings listed in table 1). We do not know whether these differences are associated with problems with the structure of the models or uncertainties associated with observed $l = 2$ mode frequencies. If the discrepancy in small spacing is due to the models then the problem is probably located in the deep interior where the small spacings are most sensitive.

The current asteroseismic data are able to distinguish the different models of convective core overshoot. We conclude that with asteroseismic data at similar levels of accuracy or better, which is of the order of ±0.5 µHz, we should be able to distinguish the effects of core overshoot in on other stars.

## 7. SUMMARY AND CONCLUSIONS

All our conclusions depend on the validity of the asteroseismic observations, specifically those of reduction scenario A from Bedding et al. (2010). Of the two reduction scenarios presented by Bedding et al. we found that their scenario B and subsets thereof, yields inconsistent results with regard to mass, mixing length parameter, age, and composition.

We found that our results are relatively insensitive to whether or not the HR diagram location and the observed mass constraints are included. That is to say, the asteroseismic data alone are constraining the models well enough to match Procyon's HR diagram location and mass.

Comparing several different formulations of core overshoot, we find that the conventional approach in which the outer boundary of chemical mixing is extended between 1.0 and 1.5 pressure scale heights, i.e., core overshoot $\beta = 1.0$ to 1.5, yields the most probable model fits to Procyon's observations. We conclude that the extra effort to implement and use the Kuhfuss nonlocal MLT may not be worth the trouble because according to the Bayesian evidence (a probability measure) it yields poorer model fits to the observations.

The Bayesian evidences also suggest that the overshoot region is not adiabatic but retains the original radiative temperature gradient. Basically, for Procyon we find a more diffusive process for the overshoot than the penetrative overshoot predicted by Zahn (1991).

We conclude that some process, which could be overshoot but also could be some other physical process that yields similar deep interior perturbations to the standard structure, is occurring in the region surrounding the convective core of Procyon. Indeed, the fact that the Kuhfuss models, which are based on a more detailed model of convection, do not appear to preform well for Procyon may be a hint that something other than convective core overshoot is taking place in Procyon's deep interior.

The age of our most probable models increases by 60% (from 1.7 Gyr to 2.7 Gyr) when convective core overshoot ($\beta = 1.0$) is included. Core overshoot mixes fresh hydrogen into the nuclear burning core. This extends the core hydrogen burning phase and the time to turnoff. Because star cluster dating depends on the luminosity and age at turnoff and because the luminosity and age at turnoff themselves depend on the amount of core overshoot, it is important to know if the amount of overshoot we see in Procyon is typical of all stars with convective cores. Dogan et al. (2010) derive an age of 1.83 Gyr using scenario A seismic data, which agrees with our non-overshoot result (1.85±0.02 Gyr). The slightly older ages for the core overshoot models is not a serious problem for Procyon because there are no independent measures of Procyon's age with which to compare. Needless to say, increasing the ages of other stars by this amount could pose serious challenges in other fields that rely on stellar isochrones.

The ambiguity in Procyon's evolutionary phase (Guenther & Demarque 1993), which could be (a) just before core hydrogen exhaustion, (b) at core hydrogen exhaustion, or (c) at the start of core contraction and hydrogen shell burning, is now removed. All our best fitting models show Procyon to be pre-core hydrogen exhaustion. This is further confirmed by the observed oscillation spectrum which does not show any bumped modes, i.e., modes whose frequencies are perturbed from the regular sequence of spacings between adjacent modes (Osaki 1975). Bumped modes are found is stars with contracted cores, i.e., stars that have evolved past core hydrogen exhaustion.

We find that the mass fraction (radius fraction) of the convective core increases from 0.07 (0.06) to 0.12 (0.07) when going from models with no overshoot to models with core overshoot $\beta = 1.0$. Chaboyer et al. (1999) obtained comparable convective core sizes, ranging from mass fraction 0.07 to 0.08 for no overshoot to 0.9 for core overshoot $\beta = 0.10$.

Most of the fundamental parameters of our best fitting models do not depend on the amount of overshoot used in the models (see Table 4). In particular, the mass, the composition, and the mixing length parameter of the most probable models do not change when overshoot is included to the model calculation. The mass and composition agree with observations within the uncertainties. And the mixing length parameter is nearly identical to the standard solar model tuned value ($\alpha = 1.8$), again, independent of the amount of overshoot included in the model.

In addition, to studying overshoot in the convective core, we also looked into the effects of the diffusion of helium and metals out of the envelope. We wanted to confirm that our Bayesian code that takes into account surface effects is unaffected by this surface effect and we wanted to test whether or not diffusion improves the models. Regarding the Bayesian code itself, it behaved as expected. The most probable models, as determined by our Bayesian code, that do include diffusion of helium and metals have nearly identical mass, age, $\alpha$, and convective core size when compared to the most probable models that do not. Furthermore, the evidences for the model subgrids peaked in the same range of $\beta$, i.e., from 1.0 to 1.5. The surface features of the most probable models are, though, not identical. And, in fact, the evidences for the model subgrids that include diffusion are higher than the evidences for the corresponding subgrids that do not include diffusion.

Despite these positive results, we are not yet ready to conclude that diffusion (gravitational settling as modeled here) is the only nonstandard physical process operating in Procyon's envelope. As we discussed in Section 3, the computation of diffusion in models with thin convective envelopes is difficult because the predicted rates are so high that helium and metals can be abruptly and totally drained from the convective envelope in a few evolutionary time steps (Morel and Thévenin 2002). When this happens our evolution code becomes unstable, being unable to fix the location of the base of the convection zone. To deal with this situation we had to turn diffusion off in the stellar evolution code whenever the convective envelope got too thin. But our ad-hoc fix to modeling diffusion is not the only issue we have with our modeling of the surface layers. We did not include convective envelope overshoot in our models, which may also be important. Like diffusion, convective envelope overshoot perturbs the higher *p*-mode frequencies but unlike diffusion, convective envelope overshoot also perturbs the lower

*p*-mode frequencies because it indirectly forces the star to burn more of its core hydrogen fuel to reach the same HR diagram position. It should be possible to distinguish the two effects using the existing asteroseismic data and extending the model grids to include a range of envelope overshoot values. This, though, will require an order of magnitude more computational resources.

The most probable model without diffusion and without core overshoot has $Z_s/X_s$ = 0.020±0.002 (see Table 4), which, although on the low side, is within uncertainties of the observed abundance (see Table 1, section 2.1) $Z_s/X_s$ = 0.0245±0.003. The most probable model without diffusion but with overshoot $\beta$ = 1.0 has $Z_s/X_s$ = 0.030±0.005, which is higher than the observed value but within the uncertainties. When diffusion and overshoot $\beta$ = 1.0 are included $Z/X$ returns closer to the solar value, $Z_s/X_s$ = 0.022±0.002. The uncertainties in the observed $Z_s/X_s$ are too large to definitively rule out any of the most probable models. Regardless, the trend is clear and if the uncertainties in $Z_s/X_s$ were smaller, then diffusion would be required in models with large values of overshoot in order to reproduce the observed $Z_s/X_s$.

The amount of overshoot found in other studies for other stars varies but is less than the amount we find for Procyon. Mermilliod & Maeder (1986) examining the turnoff location of young cluster stars, with masses between 1.25 $M_\odot$ and 9 $M_\odot$, find that models with overshoot between 0.2 $H_P$ and 0.4 $H_P$ fit the best. Similar, Demarque et al. (1994) conclude that core overshoot of ~0.23 $H_P$ is required to reproduce the morphology of the gap located near turn-off in the color-magnitude diagram for NGC 2420. Stothers and Chin (1991) show that modern opacities (Iglesias & Rogers 1991 compared to the older Cox & Stewart 1970 opacities) eliminate the need to include large amounts of overshoot in the models to explain the discrepancies previously noted in more massive stars, including the mass-luminosity relation for Cepheids. They conclude that the amount of overshoot required in massive stars is less than 0.2 $H_P$. From eclipsing binaries in the mass range 2.0 $M_\odot$ to 3.5 $M_\odot$ evolving off the main sequence, Ribas et al 2000 find overshoot values of 0.25±0.05 $H_P$ with evidence for overshoot of ~0.1 $H_P$ for stars of mass similar to Procyon. Guenther & Demarque (1993) also required little core overshoot in their earlier modeling of Procyon. Using asteroseismic data, Dupret et al. 2004 found that for β Cep star HD129929, a star with mass ~10$M_\odot$, the best fitting models have core overshoot of ~0.1 $H_P$, rejecting entirely values greater than 0.2 $H_P$. Deheuvels and Michel (2011) used CoRoT asteroseismic data for a solar-like star and the existence of a bumped mode to constrain the amount of overshoot to low values < 0.2 $H_p$. Silva Aguirre et al. (2013) modeled two Kepler stars and found a variety of different model fits to these stars but all with low values of overshoot.

  Although the cited studies do not replicate exactly the constitutive physics of our models, nor are the studies focused on stars of Procyon's mass and evolutionary phase, the fact remains, the amount of overshoot we require in our models to match the observed frequencies is greater than that found by others for other types of stars. Physically, we are asking convective motions surrounding the core of Procyon to overshoot by a significant amount. It seems, maybe more realistic, to imagine that other physical processes like circulation or turbulence induced by rotational shears are mixing the region around Procyon's core. Possibly relevant is the binary nature of the Procyon system. The progenitor of the white dwarf companion may have originated with a mass several times that of the present Procyon A and lost most of its mass during evolution. A more massive companion would also have meant a smaller orbit, and a phase of tidal interaction between Procyon's two components, resulting in internal mixing in the interior of what is now Procyon A. Regardless of the cause, we find evidence that there is some process occurring in the core of Procyon that is altering the structure in a way that looks like the effects of core overshoot.

  Further progress on Procyon will require more accurate data obtained from much longer asteroseismic observational runs to confirm the frequencies and mode identifications of the lowest frequency modes and the $l = 2$ modes and, if possible, to try to resolve rotational splittings that could provide evidence for circulation or rotational mixing. The next step for us will be to carry out similar studies using the Kepler asteroseismic data archive to see if there is evidence for overshoot (or overshoot like behavior) in other stars similar to Procyon.

DBG acknowledges the funding support of the Natural Sciences and Engineering Research Council of Canada. MG acknowledges the support of the Vanier Canada Graduate Scholarship. The authors thank Terry Girard for providing advance information on the latest HST measurements of Procyon and undergraduate student Kieran MacLeod for his help with the preliminary test runs.